\documentclass[12pt,letterpaper]{article}

\usepackage[T1]{fontenc}
\usepackage[utf8]{inputenc}
\usepackage{microtype}

\usepackage[margin=1in,headheight=15pt]{geometry}

\usepackage{setspace}
\usepackage{amsmath,amssymb}

\usepackage{booktabs}
\usepackage{longtable}
\usepackage{multirow}
\usepackage{tabularx}
\usepackage{array}
\usepackage{pdflscape}
\usepackage{ragged2e}       

\usepackage{enumitem}

\usepackage{url}
\usepackage[hidelinks,breaklinks=true]{hyperref}
\usepackage{xurl}

\usepackage{graphicx}

\usepackage{xcolor}

\usepackage{titlesec}
\titleformat{\section}{\large\bfseries}{\thesection.}{0.5em}{}
\titlespacing*{\section}{0pt}{18pt}{8pt}

\titleformat{\subsection}{\normalsize\bfseries}{\thesubsection}{0.5em}{}
\titlespacing*{\subsection}{0pt}{14pt}{6pt}

\titleformat{\subsubsection}{\normalsize\itshape}{\thesubsubsection}{0.5em}{}
\titlespacing*{\subsubsection}{0pt}{12pt}{4pt}

\usepackage{fancyhdr}
\usepackage{caption}
\usepackage{cite}

\begin{document}

\begin{titlepage}
\thispagestyle{empty}
\centering
\vspace*{2cm}
{\LARGE\bfseries WHO GOVERNS THE MACHINE?\par}
\vspace{0.5cm}
{\large A Machine Identity Governance Taxonomy (MIGT) for AI Systems\\
Operating Across Enterprise and Geopolitical Boundaries\par}
\vspace{1.5cm}
{\large Andrew Kurtz, CISSP\\
\small \href{mailto:andrewkurtz@acm.org}{andrewkurtz@acm.org}\par}
\vspace{0.5cm}
{\large Klaudia Krawiecka, PhD\\
\small \href{mailto:kkrawiecka@acm.org}{kkrawiecka@acm.org}\par}
\vspace{1cm}
{\normalsize \today\par}
\end{titlepage}


\begin{abstract}
The governance of artificial intelligence has a blind spot: the machine identities that AI systems use
to act. AI agents, service accounts, API tokens, and automated workflows now outnumber human
identities in enterprise environments by ratios exceeding 80 to 1 \cite{ref1}, yet no integrated framework
exists to govern them. The consequences are measurable: a single ungoverned automated agent
produced \$5.4 to \$10 billion in losses in the 2024 CrowdStrike outage, and nation-state actors
including Silk Typhoon and Salt Typhoon have operationalized ungoverned machine credentials as
primary vectors for espionage against critical infrastructure. At the same time, three major
jurisdictions-the European Union, the United States, and China-are developing fundamentally
incompatible AI governance frameworks, creating cross-jurisdictional conflicts that no existing
enterprise governance program is equipped to navigate.

There are four original contributions to the emerging field of AI identity governance within. First, it
proposes the AI-Identity Risk Taxonomy (AIRT), a comprehensive enumeration of risks arising
specifically at the intersection of AI systems and enterprise identity governance, comprising 37 risk
sub-categories across eight domains, each grounded in documented incidents, regulatory
recognition, practitioner prevalence data, and threat intelligence. Second, it introduces the Machine
Identity Governance Taxonomy (MIGT), an integrated governance framework organized across six
domains that simultaneously addresses the technical governance gap, the regulatory compliance
gap, and the cross-jurisdictional coordination gap that existing frameworks address only in
isolation. Third, it presents the foreign state actor threat model for enterprise identity governance,
establishing that Silk Typhoon, Salt Typhoon, Volt Typhoon, and North Korean AI-enhanced
identity fraud operations have already operationalized AI identity vulnerabilities as active attack
vectors requiring governance responses that no existing IAM framework provides. Fourth, it
proposes the cross-jurisdictional regulatory alignment structure mapping enterprise AI identity
governance obligations under EU, U.S., and Chinese frameworks simultaneously, identifying
irreconcilable conflicts and providing a governance mechanism for managing them at the enterprise
program level. A four-phase implementation roadmap translates the MIGT's six governance
domains into actionable enterprise programs. The paper is addressed to policy makers, regulators,
and practitioners operating at the intersection of AI governance, cybersecurity, and geopolitical risk.

\end{abstract}

\section*{Glossary of Key Terms}

The following terms are used throughout the paper. Definitions are provided to support readers from
policy, regulatory, and practitioner backgrounds who may not be familiar with all domains
addressed.

\begin{description}

\item[\textbf{Agentic AI.}] An artificial intelligence system capable of autonomous perception,
decision-making, and action within an environment, often executing multi-step tasks without
continuous human direction. Agentic AI systems differ from earlier generative AI systems in that
they act, not merely respond. For the purposes of this paper, that distinction carries specific
governance weight: agentic AI systems acquire and exercise access rights, consume and produce
enterprise data, invoke tools and external services, delegate authority to subordinate agents, and
accumulate de facto privileges across systems in ways that no prior generation of software required
identity governance frameworks to address.

The multidimensional operational profile of an agentic AI system comprises four analytically
distinct capabilities, each of which creates a distinct identity governance surface. First,
\emph{tool invocation}: agentic AI systems call external APIs, execute code, query databases,
trigger workflows, and interact with enterprise applications through programmatic interfaces that
carry real access authority. Each tool invocation represents an authenticated action performed
under the agent's identity and within the scope of its provisioned credentials, making tool usage a
primary vector for both privilege abuse and exfiltration. Second, \emph{multi-step autonomous
execution}: unlike a generative AI system that responds to a single prompt and terminates, an
agentic AI system plans and executes sequences of actions across time, accumulating context,
state, and effective access across each step without necessarily returning to human oversight
between them. The governance challenge this creates is not merely one of endpoint access control
but of behavioral trajectory: the aggregate effect of individually authorized steps may produce
outcomes that no single authorization decision contemplated or approved. Third,
\emph{agent-to-agent delegation}: agentic AI systems increasingly operate within multi-agent
architectures, spawning subagents, delegating subtasks to specialized agents, and passing
instructions and credentials through agent chains whose trust relationships are rarely
cryptographically verified and whose accountability structures are rarely explicitly assigned. Each
delegation boundary is a potential accountability gap and a potential privilege escalation surface.
Fourth, \emph{environmental interaction and memory}: agentic AI systems read from and write to
persistent storage, ingest content from documents, emails, databases, and web pages, and in some
architectures maintain memory across sessions. Each of these interactions creates a pathway for
indirect prompt injection, unauthorized data aggregation, and cross-session privilege accumulation
that static access governance models are structurally unable to detect.

Taken together, these four capabilities mean that an agentic AI system is not a passive
computational resource but an active, identity-bearing enterprise actor whose access footprint,
behavioral trajectory, and accountability structure require governance disciplines equivalent to, and
in several respects more demanding than, those applied to privileged human users.

\item[\textbf{AIRT (AI-Identity Risk Taxonomy).}] The AI-Identity Risk Taxonomy proposed in
Section~5. A two-level classification of 37 risk sub-categories across eight domains, comprising
comprehensive enumeration of risks arising specifically at the intersection of AI systems and
enterprise identity governance.

\item[\textbf{API Key.}] An Application Programming Interface key. A static credential used to
authenticate a software system or AI agent to an external service. API keys are a primary target
of state-sponsored credential theft operations because they provide persistent, often broad access
to enterprise systems and AI infrastructure.

\item[\textbf{ARIA (Agent Relationship-Based Identity and Authorization).}] A governance model that
records every agent-to-agent delegation as a distinct, cryptographically verifiable relationship in a
governance graph, enabling fine-grained accountability across complex multi-agent workflows.

\item[\textbf{ATF (Agentic Trust Framework).}] The Agentic Trust Framework published by the Cloud
Security Alliance in February 2026. A zero-trust governance framework for AI agents that
introduces a four-maturity-level model in which agent autonomy is earned through demonstrated
trustworthiness rather than granted as a binary condition.

\item[\textbf{CAC (Cyberspace Administration of China).}] The primary Chinese government regulatory
body responsible for internet content governance, cybersecurity oversight, and AI algorithm
regulation, including administration of China's algorithm filing regime.

\item[\textbf{CSL (Cybersecurity Law).}] China's foundational cybersecurity statute, originally enacted
in 2017 and significantly amended effective January 1, 2026, to incorporate dedicated AI
compliance provisions and expanded extraterritorial reach.

\item[\textbf{DID (Decentralized Identifier).}] A W3C standard for a globally unique, cryptographically
verifiable identifier that does not require a centralized registration authority. DIDs are proposed
as a foundation for AI agent identity because they enable agents to prove their identity and
provenance to relying systems without dependence on a single identity provider.

\item[\textbf{EU AI Act.}] The European Union's Artificial Intelligence Act, which entered into force on
August~1, 2024 and is the world's first comprehensive binding AI regulation. It classifies AI
systems by risk level and imposes obligations proportional to risk, with the most significant
high-risk AI system requirements taking effect on August~2, 2026.

\item[\textbf{GPAI (General-Purpose AI Model).}] Under the EU AI Act, a foundation model trained on
broad data at scale that can perform a wide range of tasks. GPAI models are subject to specific
obligations under Articles~53 and~55 of the Act, including technical documentation, copyright
compliance, and training data transparency requirements.

\item[\textbf{IAM (Identity and Access Management).}] The discipline and set of technical systems
responsible for managing digital identities and controlling access to enterprise resources.
Traditional IAM frameworks were designed primarily for human user identities and are
structurally inadequate for the governance of AI agent identities.

\item[\textbf{JIT (Just-in-Time) Provisioning.}] An access governance approach in which access rights
are granted at the moment they are needed for a specific task and automatically revoked upon task
completion, rather than being pre-assigned as standing entitlements. JIT provisioning eliminates
standing privilege and directly addresses the Dynamic Access Requirement Mismatch risk in AIRT
Domain~II.

\item[\textbf{MCP (Model Context Protocol).}] An emerging standard for AI agent-to-application
connectivity that enables AI agents to retrieve data, trigger workflows, and act within enterprise
systems. MCP connections carry real access authority and represent an emerging identity
governance surface addressed in Section~8.4 as a future research direction.

\item[\textbf{A2A (Agent-to-Agent Protocol).}] An open protocol developed by Google and published in
April 2025 that enables communication, task delegation, and capability discovery between AI
agents operating across heterogeneous frameworks and organizational boundaries. Where MCP
governs how an individual agent connects to tools and data sources, A2A governs how agents
find, authenticate, and coordinate with one another. Each agent publishes an \emph{Agent Card},
a structured JSON manifest declaring its capabilities, supported modalities, and authentication
requirements, which peer agents and orchestration systems use to determine whether and how to
delegate tasks. For the purposes of this paper, A2A is a primary emerging identity governance
surface: agent-to-agent delegation relationships established through A2A carry real access
authority, cross organizational and trust boundaries, and currently lack the cryptographically
verifiable accountability structures that the MIGT's Domain~II and Domain~IV controls require.

\item[\textbf{MIGT (Machine Identity Governance Taxonomy).}] The Machine Identity Governance
Taxonomy proposed in Section~6. A six-domain integrated governance framework addressing
the technical governance gap, the regulatory compliance gap, and the cross-jurisdictional
coordination gap simultaneously. The MIGT provides governance capabilities, controls, and
accountability mechanisms for enterprise AI identity programs.

\item[\textbf{NCCoE (National Cybersecurity Center of Excellence).}] A division of NIST that develops
practical, standards-based cybersecurity solutions. The NCCoE initiated a project specifically
addressing software and AI agent identity and authorization in 2026, signaling that authoritative
U.S.\ standards for AI agent identity governance are forthcoming.

\item[\textbf{NHI (Non-Human Identity).}] Any digital identity that is not associated with a human user.
NHIs include AI agents, service accounts, API tokens, OAuth clients, automated workflows, and
device identities. NHIs now outnumber human identities in enterprise environments by ratios as
high as 144 to 1.

\item[\textbf{NIST AI RMF.}] The NIST Artificial Intelligence Risk Management Framework (AI RMF
1.0), published in 2023. The de facto operational standard for AI governance in the United
States, establishing the GOVERN, MAP, MEASURE, and MANAGE functions. The AI RMF
provides governance function structure but does not enumerate identity-specific AI risks-the
gap the AIRT is designed to fill.

\item[\textbf{PAM (Privileged Access Management).}] A category of identity security tools and practices
focused on managing, monitoring, and auditing access by privileged accounts, including
administrator accounts, service accounts, and high-privilege machine identities. PAM provider
credentials have been identified as a primary target of the Silk Typhoon state-sponsored
espionage campaign.

\item[\textbf{SPIFFE / SPIRE.}] Secure Production Identity Framework for Everyone (SPIFFE) is a CNCF
graduated standard for cryptographic workload identity. SPIRE is its reference implementation.
SPIFFE provides cryptographically verifiable identities tied to workloads rather than people,
making it the recommended cryptographic identity standard for AI agent deployments in
enterprise environments.

\item[\textbf{ZSP (Zero Standing Privilege).}] An access governance principle under which AI agents and
other machine identities never possess a credential or access right at rest. Access is granted
just-in-time for specific tasks and automatically revoked upon completion. ZSP is the target
architecture for MIGT Domain~II and directly addresses the Static Credential Persistence risk
in AIRT Domain~I.

\end{description}

\pagestyle{plain}
\tableofcontents
\newpage
\pagestyle{fancy}

\section{Introduction: The Governance Vacuum at the Intersection of AI and Identity}

In 2026, most organizations cannot answer three basic questions about their AI systems: What can
they access? Who is responsible when they cause harm? And does governing them in one country
create liability in another?

These are not edge-case questions. They are the foundational questions of enterprise risk
management, namely the same questions that identity and access management (IAM) programs have
answered for human identities for three decades. The answer to all three, for most organizations in
2026, is: we do not know. That is the governance vacuum that is addressed.

At the operational level, the governance vacuum manifests in three questions that practitioners
working directly with agentic AI deployments report being asked with increasing urgency: What
tools should an agent be given to complete a task? What data is it permitted to process in doing so?
And does the combination of tool access and data access remain within the authorized scope of the
task it was assigned (i.e., the original intent)? These are not abstract policy questions. They are the
runtime access governance decisions that determine whether an AI agent operates within its
sanctioned boundaries or accumulates capability and data exposure that no single authorization
decision contemplated. The absence of governance frameworks capable of answering them at the
speed and scale at which agentic AI systems operate is the operational expression of the governance
vacuum this paper addresses. Existing IAM frameworks were not designed to evaluate
tool-permission combinations dynamically, to enforce task-scoped data access boundaries at
machine speed, or to detect when the aggregate effect of individually authorized tool invocations
exceeds the scope of the task that justified them. The result is that organizations are making these
decisions implicitly, through the access grants they provision at deployment time, without the
governance infrastructure to verify that those grants remain appropriate as agent behavior evolves
across tasks, sessions, and system interactions.

The governance vacuum is not primarily a technical failure. It is a framework failure, specifically a
failure to extend the disciplined, accountable governance structures that organizations have built for
human identities to the machine identities that now outnumber them by more than 80 to 1
\cite{ref1}\cite{ref2}.

That framework failure has a structural dimension that distinguishes it from ordinary governance
lag and that makes it resistant to the incremental remediation that has historically closed gaps
between emerging technology and governance practice. Agentic AI is among the fastest-developing
capability domains in the history of enterprise technology: the transition from generative AI systems
that respond to agentic AI systems that act, plan, delegate, and accumulate access has occurred
within a timeframe measured in months rather than the years or decades over which prior technology
transitions allowed governance frameworks to mature. The consequence is that frameworks
developed to address agentic AI governance risks become narrow relative to current capability and
begin to go out of date before they achieve practitioner adoption, creating a persistent velocity
asymmetry between the frontier of agentic AI deployment and the frontier of agentic AI governance.
No single framework, however carefully designed, can remain current across all dimensions of a
domain evolving at this speed. What is required is not a point-in-time framework but a
comprehensive, multidimensional governance structure whose domains are stable enough to provide
durable organizational anchors while remaining extensible as specific risk categories, technical
standards, and regulatory obligations evolve beneath them. The MIGT is designed to provide
exactly that structure: a framework whose six governance domains address the persistent structural
characteristics of the AI identity governance problem-the identity lifecycle, the authentication
architecture, the access governance model, the accountability attribution mechanism, the supply
chain integrity requirement, and the cross-jurisdictional coordination challenge-rather than the
specific technical manifestations of those characteristics at any single moment in time.

The framework failure does not exist in isolation. Compounding it is a regulatory failure of
equivalent structural depth: an expanding body of AI-specific law is being written by legislators and
regulators who, in many cases, do not understand the specific mechanics of identity and access
governance, and an expanding body of IAM practice is being built by practitioners not adequately
tracking the regulatory obligations their programs must satisfy \cite{ref4}\cite{ref7}. The regulatory failure
compounds the framework velocity problem: regulatory instruments, which by their nature require
extended drafting, consultation, and legislative processes, are structurally unable to track a
technology domain evolving at the pace agentic AI is evolving. The result is regulation that
addresses the AI capabilities of the drafting period rather than the deployment period, creating
compliance frameworks that are simultaneously over-specified in areas where the technology has
moved on and under-specified in areas-including agent tool governance, task-scoped data access,
and multi-agent accountability attribution-where current deployment practice has outpaced
regulatory imagination.

Where the framework failure and the regulatory failure operate primarily within organizational and
national boundaries respectively, the third dimension of the governance vacuum is irreducibly
global: a coordination failure in which every major jurisdiction is developing its own AI governance
framework in relative isolation, creating a patchwork of potentially incompatible obligations that no
existing framework has mapped against the realities of enterprise identity programs
\cite{ref12}\cite{ref30}.

This paper argues that solving the AI identity governance problem requires simultaneously
addressing all three failures: the framework gap, the regulatory gap, and the coordination gap.
Addressing any one in isolation produces incomplete solutions. A technically rigorous AI identity
governance framework that ignores regulatory obligations will fail compliance requirements. A
regulatory compliance program that lacks technical framework grounding will fail operational
security requirements. And both will fail when organizations encounter the cross-jurisdictional
conflicts that are already materializing as AI systems operate across borders governed by
fundamentally incompatible legal philosophies \cite{ref31}\cite{ref32}.

\subsection{The Scope and Urgency of the Problem}

The scale of the AI identity governance problem is empirically documented and growing. The Entro
Labs NHI and Secrets Risk Report for H1 2025 found that non-human identities grew 44\%
year-over-year and now outnumber human identities at a ratio of 144 to 1, up from 92 to 1 in H1
2024 \cite{ref45}. Gyure and Johnson \cite{ref2} document machine identity growth from approximately
50,000 per enterprise in 2021 to 250,000 by 2025, a 400\% increase in four years, while governance
capabilities have not kept pace. IBM Security researchers have identified that IAM teams are
typically responsible for only 44\% of an organization's machine identities, leaving the majority
essentially ungoverned \cite{ref3}. The Delinea 2026 security forecast states plainly that if 2025 was the
year organizations embraced AI, 2026 will be the year they lose track of it \cite{ref4}.

The consequences of this governance gap are not hypothetical. The 2024 CrowdStrike outage,
which was caused by a single automated software update agent operating across enterprise
environments globally without adequate governance controls, produced an estimated \$5.4 to \$10
billion in economic losses \cite{ref84}\cite{ref85} and demonstrated how a single ungoverned machine
identity can cause cascading failures at planetary scale. The Silk Typhoon espionage campaign,
active since 2024, has used stolen API keys and other machine credentials to exfiltrate data of
interest to the Chinese government, demonstrating that foreign state actors have already identified
ungoverned machine identities as a primary attack vector \cite{ref5}\cite{ref58}. DeepSeek's alleged
exfiltration of OpenAI training data through developer API accounts in late 2024 illustrates the AI
supply chain dimension of this risk, where machine identities serve as both the attack vector and
the target \cite{ref5}.

The State of AI Agent Security 2026 report, based on survey data from over 900 executives and
technical practitioners, found that 88\% of organizations confirmed or suspected AI agent security
incidents in the preceding year, yet only 22\% of teams treat AI agents as independent,
identity-bearing entities, and 45.6\% still rely on shared API keys for agent-to-agent authentication
\cite{ref89}. The governance gap documented in Section~2 is not a theoretical projection: it is an
operational crisis already producing measurable, attributable harm.

\subsection{Why Existing Frameworks Are Insufficient: Five Structural Gaps}

The insufficiency of existing frameworks is structural rather than incremental. These gaps have been
identified across multiple bodies of literature reviewed in Section~3, including the AI agent security
survey literature \cite{ref34}\cite{ref37}, the agentic AI governance literature \cite{ref28}\cite{ref29}\cite{ref38}, and the
IAM standards literature \cite{ref26}\cite{ref27}\cite{ref7}.

\textbf{Gap 1, The Identity Continuum Problem.} Current IAM frameworks treat human and machine
identities as binary categories with separate governance approaches. Gyure and Johnson \cite{ref2}
demonstrate that this binary categorization fails to address the nuanced reality of modern enterprise
identities, which exist on a spectrum from fully human-controlled to fully autonomous. An AI agent
acting on behalf of a human, executing their instructions, using their delegated permissions, but
making autonomous decisions in execution, is neither fully human nor fully machine for governance
purposes. Frameworks built on binary categorization cannot govern the spectrum.

\textbf{Gap 2, The Dynamic Access Problem.} Role-based access control assumes that roles can be
defined and that access requirements are stable. Agentic AI systems discover their access
requirements dynamically, at runtime, as a function of the tasks they are executing
\cite{ref7}\cite{ref27}. The OpenID Foundation \cite{ref27} has documented that current OAuth 2.0 and
SAML standards lack native support for the machine-speed reauthentication and multi-party
delegation that agentic AI systems require.

The dynamic access problem is compounded by a structural characteristic of agentic AI that has no
precedent in human or traditional service identity governance: the capacity to autonomously
decompose a task into parallel sub-tasks and spawn multiple subordinate agents to execute them
simultaneously. A human employee working on a complex task proceeds sequentially, their access
requirements unfolding in a predictable, auditable order that governance frameworks can anticipate
and bound. A service identity executes a defined workflow whose access requirements are known at
design time and do not change between invocations. An agentic AI system does neither. It may
determine at runtime that a task is most efficiently accomplished by spawning three sub-agents in
parallel, each requiring distinct access grants to distinct systems, each potentially spawning further
sub-agents of their own, producing an access demand profile that is neither predictable at
provisioning time nor bounded by any fixed workflow definition.

This is the governance consequence of the non-deterministic nature of agentic AI: unlike a
conventional computer program, whose execution path can be represented as a flow chart to which
the program commits at design time, an agentic AI system's execution path is an emergent property
of its reasoning process, the tools available to it, the data it encounters, and the intermediate results
it produces. You can no longer draw the flow chart. The access requirements of step four depend on
what the agent decided at step three, which depended on what it found at step two, none of which
was determined, or determinable, when the agent's credentials were provisioned. Traditional
role-based access control is built on the assumption that the flow chart exists and is known in
advance: roles are defined because the tasks that require them are defined. When the task structure
itself is non-deterministic, the role structure that governance frameworks depend on cannot be
specified in advance, and the governance choice between over-provisioning, which creates risk, and
under-provisioning, which breaks functionality, is not merely a false dilemma created by applying a
static framework to a dynamic problem, as prior analyses have characterized it \cite{ref7}\cite{ref27}: it
is an unavoidable consequence of applying any provisioning-time governance model to a system
whose access requirements are determined at reasoning time. The multi-agent spawning dimension
extends this problem across an additional axis: each sub-agent represents an independent identity
with its own access grants, its own behavioral trajectory, and its own accountability record, yet the
governance frameworks that must bound their collective behavior were designed for identities that
act one at a time, in sequence, along paths that a human architect drew before the system ran.

The governance choice between over-provisioning (which creates risk) and under-provisioning
(which breaks functionality) is a false dilemma created by applying a static framework to a dynamic
problem.

\textbf{Gap 3, The Accountability Attribution Problem.} When a human employee causes harm,
accountability attribution is straightforward. When an AI agent causes harm, accountability is
distributed across the agent's developers, the organization that deployed it, the prompt or instruction
that triggered the harmful behavior, and the access permissions that made the harm possible
\cite{ref8}.

That distribution of accountability reflects a fundamental ontological shift in what an AI agent is,
relative to every prior category of enterprise software. Traditional software systems, including
earlier generations of automation and robotic process automation, were digital extensions of human
actions: a human defined the workflow, a human authorized the execution, and the software
performed exactly and only what the human specified. Accountability attribution was also
straightforward because the causal chain from human decision to system action was direct,
auditable, and complete. The human who designed the workflow owned its consequences because
the workflow owned nothing of its own: no judgment, no initiative, no capacity to determine that
the situation called for something other than what was specified.

Agentic AI systems break this causal chain. They are no longer a digital extension of human
actions. They are autonomous `digital workers': entities that receive a goal rather than a procedure,
determine their own approach to achieving it, select and invoke the tools they judge appropriate,
spawn subordinate agents when they assess that parallelization would be efficient, and produce
outcomes that their human principals neither specified nor, in many cases, could have anticipated.
The accountability question this creates is not merely one of distributing blame across a longer
chain of human decisions. It is one of attributing responsibility for outcomes that no human decision
fully determined, because the decisions that produced them were made, at least in part, by the agent
itself. This is a category of accountability problem that enterprise governance frameworks have
never been required to address, because no prior category of enterprise software was capable of
making decisions in the relevant sense.

The practical governance consequence is that the accountability structures organizations have
built-designated approvers, audit trails of human authorizations, change management processes,
incident ownership assignment-all presuppose that a human decision sits at the origin of every
consequential system action. When the consequential decision is made by an agent reasoning
autonomously toward a goal, those structures do not fail at the edges: they fail at the foundation.
There is no approver to identify because no human approved the specific action. There is no
authorization record because the action was not authorized as a discrete step but emerged from the
agent's autonomous execution of a broader mandate. The audit trail records what the agent did but
not why it decided to do it, and in a non-deterministic system whose reasoning process is not fully
interpretable, the why may not be fully reconstructable even in principle.

Shavit et al.\ \cite{ref38} propose unique agent identifiers as one mechanism for accountability
attribution, but note that no standardized system yet exists. Chan et al.\ \cite{ref8} establish that
diffuse accountability is the condition under which governance failures proliferate without
consequence. Holgersson et al.\ \cite{ref39} extend this analysis through a principal-agent framework,
arguing that the complexity of multi-agent systems creates unpredictabilities that existing
governance frameworks are not designed to address.

\textbf{Gap 4, The Foreign Influence and State Actor Problem.} Traditional IAM threat models focus
on external attackers and insider threats. AI identity governance must add a third category: foreign
state actors who use AI systems as instruments of espionage, influence, and economic warfare
\cite{ref5}\cite{ref9}\cite{ref10}\cite{ref23}. The Silk Typhoon campaign's use of stolen API keys
\cite{ref58}\cite{ref59}, North Korea's deployment of AI-enhanced operatives \cite{ref10}, and documented
use of AI-generated content for foreign influence operations \cite{ref11} demonstrate that foreign state
exploitation of AI identity governance gaps is a present operational reality. No existing IAM
framework has a foreign state actor threat model.

\textbf{Gap 5, The Regulatory Fragmentation Problem.} Organizations operating globally face AI
governance obligations under the EU AI Act, U.S.\ executive orders and sector-specific guidance,
China's Cybersecurity Law and AI governance frameworks, and a proliferating array of data
governance compliance programs, all reflecting genuinely different national philosophies and
creating cross-jurisdictional conflicts \cite{ref30}\cite{ref31}\cite{ref32}. Lim \cite{ref31} identifies regulatory
fragmentation as accelerating rather than converging between 2023 and 2025. Perboli et al.\
\cite{ref95} confirm that without coordination two specific risks materialize: regulatory arbitrage and
normative fragmentation.

\textbf{Gap 6, The Cross-Jurisdictional Agent Value Misalignment Problem.} The five structural
gaps identified above share a common assumption: that the governance failures they describe are,
in principle, correctable through better technical controls, clearer regulatory frameworks, and more
disciplined organizational practice. Gap 6 is of a different character. It arises not from the absence
of governance but from the presence of fundamentally incompatible governance philosophies that
are instantiated directly into the behavioral objectives of the AI agents those governance systems
produce.

AI agents do not arrive at enterprise environments as neutral computational instruments. They are
trained, fine-tuned, and aligned against value frameworks, safety criteria, and behavioral objectives
that reflect the normative commitments of their developers, and by extension, the regulatory,
cultural, and geopolitical environments in which those developers operate. Empirical research
confirms that large language models exhibit measurably divergent value patterns across national and
cultural contexts, reflecting the normative environments of their training data and alignment
processes. An agent developed and aligned under the EU AI Act's fundamental rights framework,
with its emphasis on human dignity, non-discrimination, and meaningful human oversight, will
instantiate a different behavioral disposition than an agent developed under China's framework,
which prioritizes social stability, state security, and content alignment with national policy
objectives, or an agent developed under the U.S.\ framework's current emphasis on capability and
competitiveness. These are not merely differences of regulatory compliance posture. They are
differences in what the agent has been trained to treat as an acceptable outcome, a permissible
action, and a legitimate objective \cite{ref95}.

The governance consequence is that when agents developed under incompatible value frameworks
interact within multi-agent architectures-which is increasingly the operational reality for global
enterprises-their behavioral objectives may be not merely different but actively conflicting. An
action that one agent's alignment framework classifies as a required safety intervention, refusing to
process certain content, escalating a decision to human oversight, declining to delegate to a
counterpart agent, may be classified by a peer agent's alignment framework as an unwarranted
obstruction of a legitimate task. The multi-agent systems literature establishes that without shared
normative frameworks, agent societies produce persistent coordination failures of precisely this
character, structural rather than incidental, and not resolvable through local optimization by any
individual agent. The ACM Computing Surveys survey of AI agent security threats confirms that
conflicting agent objectives within multi-agent architectures represent a critical and
undercharacterized knowledge gap, explicitly identifying incentive misalignment as a mechanism
for cascading failures and emergent adversarial behavior \cite{ref34}. Neither agent is malfunctioning
by its own governance criteria. Both are optimizing for objectives that their respective alignment
frameworks treat as correct. The result, from the perspective of any single governance framework,
is behavior that appears rogue, unpredictable, or adversarial, while remaining entirely coherent and
compliant from the perspective of the agent producing it.

This misalignment dynamic maps directly onto the difficulty humans have experienced in agreeing
on universal principles for AI governance. The scholarly literature documents that the EU, U.S., and
Chinese frameworks reflect genuinely incompatible internal logics rather than different technical
implementations of shared values, and that international coordination efforts have thus far failed to
resolve these incompatibilities at the human policy level. When those incompatible logics are
encoded into agent objective functions and deployed into shared multi-agent environments, the
coordination failure does not remain at the policy level: it becomes an operational reality at machine
speed, without the deliberative mechanisms that human diplomacy, however imperfect, provides.

Existing IAM frameworks have no mechanism for detecting, classifying, or governing behavioral
conflicts that arise from value misalignment rather than credential compromise or access policy
violation. The MIGT's Domain~VI Regulatory Alignment structure addresses the compliance
dimension of cross-jurisdictional divergence; the agent value misalignment problem requires an
additional governance capability-behavioral alignment verification at the agent interaction
boundary-that no current framework provides and that Section~8.4 identifies as a priority future
research direction.

\subsection{Contributions: Four Original Claims}

\textbf{Claim 1:} AI identity risk is a distinct risk category. The risks posed by AI identities,
including agentic AI systems, non-human identities, and the human-machine identity spectrum, are
not merely extensions of existing identity risk. They constitute a distinct category that requires its
own taxonomy, its own governance framework, and its own regulatory mapping. The MIT AI Risk
Repository \cite{ref13}, which classifies 1,612 AI risks, does not treat identity governance as a primary
risk domain. The peer-reviewed AI agent security surveys in ACM Computing Surveys
\cite{ref34}\cite{ref35} address agent threats without connecting them to enterprise identity governance
programs. The AIRT proposed in Section~5 is the first attempt to enumerate this risk category
comprehensively.

\textbf{Claim 2:} The governance framework gap, the regulatory gap, and the coordination gap must
be addressed simultaneously. Addressing any one in isolation produces incomplete and operationally
inadequate solutions. The agentic AI governance frameworks proposed by Pandey \cite{ref28}, Joshi
\cite{ref29}, and Shavit et al.\ \cite{ref38}, and the practitioner-oriented security framework provided by
the OWASP Top 10 for Agentic Applications \cite{ref65}, address governance broadly without
identity-specific grounding. The NIST AI RMF \cite{ref15} provides governance function structure
without identity-specific risk enumeration. Cross-jurisdictional analyses \cite{ref30}\cite{ref31} document
regulatory divergence without enterprise-level resolution mechanisms. The MIGT is designed to
address all three simultaneously: providing technical governance structure, regulatory alignment,
and cross-jurisdictional coordination within a single integrated framework.

\textbf{Claim 3:} Foreign state actors have already operationalized AI identity as an attack and
influence vector, and enterprise IAM programs have no adequate threat model for this. The evidence
reviewed in this paper, including the Silk Typhoon campaign \cite{ref5}\cite{ref58}\cite{ref59}, North Korean
AI-enhanced identity fraud \cite{ref10}\cite{ref23}, Chinese AI-generated influence operations \cite{ref11},
and the ongoing Typhoon campaign series \cite{ref60}\cite{ref61}\cite{ref62}, demonstrates that foreign state
exploitation of AI identity governance gaps is a present operational reality. An adequate AI identity
governance framework must include a foreign state actor threat model.

\textbf{Claim 4:} The absence of inter-jurisdictional coordination in AI governance is itself a
governance risk, and proposes a framework for managing it. The divergent AI governance
frameworks of the EU, U.S., and China reflect genuinely incompatible governance philosophies
that create cross-jurisdictional conflicts for every global organization deploying AI systems
\cite{ref30}\cite{ref31}\cite{ref95}. Schmitt \cite{ref96} traced the foundational architecture of this fragmented
landscape in 2021, documenting the risk of geopolitical encroachment into technical governance
domains-a warning that has fully materialized by 2026. No existing framework provides
organizations with tools for managing these conflicts systematically at the enterprise identity
governance program level.

\subsection{Paper Organization}

Section~2 establishes the empirical foundation for the governance gap, documenting the scale,
scope, and structural characteristics of the non-human identity problem. Section~3 connects this
contribution to the research tradition from which it draws, including the prior taxonomic work that
provides its methodological foundation. Section~4 analyzes the geopolitical and regulatory
landscape in detail across five major frameworks. Section~5 presents the AI-Identity Risk Taxonomy
(AIRT). Section~6 presents the Machine Identity Governance Taxonomy (MIGT). Section~7
provides implementation guidance for practitioners. Section~8 concludes with discussion of
implications and future research directions.

\section{The Empirical Foundation: Scale, Consequences, and Structural Characteristics of the Non-Human Identity Governance Problem}

This section establishes the empirical foundation for the governance gap across four dimensions:
(1) the quantitative scale of machine identity proliferation, (2) the operational consequences of
governance failure through documented incidents with measurable impact, (3) the structural
characteristics that make the gap resistant to incremental remediation, and (4) the market and
regulatory signals confirming the problem has reached a strategic inflection point.

\subsection{The Quantitative Scale of Machine Identity Proliferation}

Multiple independent research programs \cite{ref21}, \cite{ref78}, \cite{ref79}, \cite{ref90} have converged on a
consistent picture: machine identities vastly outnumber human identities, the gap is widening
rapidly, and governance capabilities are not keeping pace. The Entro Labs NHI and Secrets Risk
Report for H1 2025 found that non-human identities grew 44\% year-over-year and now outnumber
human identities at a ratio of 144 to 1, up from 92 to 1 in H1 2024 \cite{ref45}. CyberArk's 2025
Identity Security Landscape, based on a Vanson Bourne survey of 2,600 cybersecurity
decision-makers across 20 countries, confirms that machine identities vastly outnumber human
identities within organizations, driven primarily by cloud and AI, and that 68\% of respondents
report their organizations lack identity security controls for AI specifically \cite{ref78}.

The ESG Non-Human Identity Management report found that organizations on average have 20
times more non-human identities than human identities, with 52\% of organizations predicting a
further 20\% increase in the next year, and that two-thirds had experienced a successful cyberattack
resulting from compromised non-human identities \cite{ref79}. Research and Markets' February 2026
global NHI solutions report confirms that machine-to-human identity ratios exceed 17 to 1 even in
conservative enterprise estimates, with the highest-exposure environments in financial services
showing ratios as high as 96 to 1, and projects the NHI solutions market at transformational growth
through 2030 \cite{ref81}. The Veza State of Identity Security report, measuring more than 230 billion
permissions across its enterprise dataset, found that permissions classified as safe and compliant
dropped from 70\% in 2024 to 55\% in 2025, while ungoverned permissions rose from 5\% to 28\%
of total, and that 1 in 20 AWS machine identities carries full administrator privileges \cite{ref80}.
Gartner projects that by 2028, one-third of enterprise applications will include autonomous AI
agents, with the volume of agentic identities projected to exceed 45 billion by end of 2025, more
than 12 times the number of humans in the global workforce \cite{ref82}.

\subsection{Operational Consequences: Documented Incidents}

The CrowdStrike outage of July 19, 2024, caused by a single automated software update agent
deployed without adequate governance controls across 8.5 million systems simultaneously,
produced \$5.4 billion in Fortune 500 losses alone, with global estimates reaching \$10 billion
\cite{ref84}\cite{ref85}\cite{ref86}. Cybersecurity insurance policies covered only 10\% to 20\% of losses
\cite{ref84}. In May 2025, a permission was granted via a Georgia court with Delta Air Lines'
negligence suit against CrowdStrike to proceed \cite{ref87}. Tufin's analysis of the incident's lasting
impact documents that it forced a fundamental reconsideration of automated update governance
and the accountability structures required for machine identity-driven operations at enterprise
scale \cite{ref83}. The Harvard Business Review characterized the incident as a watershed moment for
machine identity governance, noting that the absence of staged rollout controls, behavioral
monitoring, and accountability assignment created the conditions for catastrophic failure at
planetary scale \cite{ref88}. Every governance control specified in MIGT Domains III and IV would
have been directly relevant to preventing or limiting the scale of this incident.

The March 2025 tj-actions supply chain compromise, in which attackers used a stolen personal
access token to inject malicious code into a widely-used GitHub Action, exfiltrating secrets from
more than 23,000 repositories, illustrates the simultaneous manifestation of AIRT Domain VI
(supply chain compromise) and Domain I (static credential persistence) \cite{ref45}. The Cloudflare
and U.S.\ Treasury incidents of 2024 followed the same pattern: under-protected API keys and
misconfigured machine identity credentials providing persistent access to critical systems without
triggering governance controls \cite{ref82}. The Entro Labs report documents that over 23.7 million
secrets were found exposed on public GitHub repositories in 2024 alone, with 57\% originating in
source code and 26\% in CI/CD workflows \cite{ref45}.

The February 2025 Marks and Spencer cyberattack, attributed to the Scattered Spider collective,
offers a further illustration of how identity governance failures across organizational boundaries
produce cascading operational harm at scale \cite{ref97}\cite{ref98}. Attackers impersonated a legitimate
M\&S employee and convinced a third-party IT provider, Tata Consultancy Services, to reset
internal credentials, gaining initial access and subsequently exfiltrating the Windows domain
controller's NTDS.dit file, which contained password hashes for every domain user including
service accounts and machine identities. Armed with these credentials, attackers moved laterally
through M\&S's corporate domain encrypted VMware ESXi infrastructure supporting e-commerce
and logistics, and forced a 46-day suspension of online ordering that produced an estimated
\pounds{}300 million in lost operating profits and a \pounds{}750 million loss in market value
\cite{ref97}. The incident illustrates the AIRT Domain V cross-organizational accountability gap, the
third-party identity governance failure the MIGT's Domain I lifecycle governance and Domain IV
accountability attribution mechanisms are specifically designed to address, and underscores that
identity governance failures in partner and vendor ecosystems are operationally indistinguishable
in their consequences from failures in the organization's own environment.

\subsection{Structural Characteristics of the Governance Gap}

Three structural features make the AI identity governance problem resistant to incremental
resolution. First, the visibility problem: organizations do not have the foundational inventory and
ownership assignment - specifically the AI Identity Registry specified in MIGT Domain I - from
which visibility flows. Without the registry, no tool can provide comprehensive visibility because no
authoritative record of what should exist is available against which actual inventory can be
compared. The CyberArk report confirms that IAM teams are responsible for only 44\% of machine
identities \cite{ref78}, and that 47\% of organizations cannot secure shadow AI usage at all
\cite{ref78}.

Second, the governance velocity mismatch: machine identities are created at machine speed,
automatically, continuously, and without governance review in most organizations. Any governance
approach relying on human-speed review perpetually lags behind the identity creation rate. Third,
the identity debt accumulation dynamic: the Veza finding that ungoverned permissions rose from
5\% to 28\% of enterprise total in a single year \cite{ref80} represents not a stable problem but an
accelerating one. Like financial debt, identity debt has compounding dynamics that existing IAM
approaches are not reversing.

\subsection{Market and Regulatory Signals: The Inflection Point}

The IAM market is projected to grow from \$19.80 billion in 2024 to \$61.74 billion by 2032, with
non-human identity governance identified as the primary growth driver \cite{ref6}. The CyberArk
report found that 88\% of respondents are under increased pressure from insurers mandating
enhanced privilege controls for machine identities, and that 90\% of global leaders cite identity
attacks as their top cybersecurity concern \cite{ref78}. Delinea \cite{ref4} characterizes 2026 as the year
when machine identities move from a technical issue to a board-level priority. The NIST CAISI AI
Agent Standards Initiative, launched February 18, 2026, and the companion NCCoE concept on AI
agent identity and authorization \cite{ref18}\cite{ref19}\cite{ref44} signal that the U.S.\ government acknowledges
no adequate standard currently exists - precisely the gap the MIGT is designed to fill.

\section{Prior Work and the Research Gap}

The research presented sits at the convergence of five distinct bodies of literature: (1) the prior
taxonomic work on non-human entity security that provides its methodological foundation, (2) AI
risk and governance frameworks, (3) identity and access management governance standards, (4)
agentic AI accountability and harm frameworks, and (5) cross-jurisdictional AI regulatory analysis.
Each body of literature makes essential contributions to the problem domain; none addresses it in
its entirety. This section surveys each, identifies its contribution and limitations relative to the
problem addressed, and constructs the cumulative case for why the MIGT and AIRT constitute a
necessary and original contribution.

\subsection{Methodological Lineage: From Non-Human Device Security to AI Identity Taxonomy}

The taxonomic methodology employed in the AIRT builds on the structured classification approach
established in Rizvi, Kurtz, Pfeffer, and Rizvi \cite{ref25}, which presented a multi-domain security
taxonomy for IoT environments at IEEE TrustCom/BigDataSE 2018. That work identified identity
and data theft, device manipulation, and data falsification as primary risk vectors in distributed
non-human entity environments, establishing the classification framework the AIRT extends to the
agentic AI domain. The present paper's contribution is not an incremental update to IoT security
taxonomy but a new application of the taxonomic method to a risk domain-AI identity
governance-that the IoT literature did not anticipate and has not addressed. Agentic AI identities
reason about their access requirements dynamically, accumulate permissions across multi-system
interactions, and operate in ways that create accountability structures with no precedent in device
security, requiring a purpose-built taxonomy rather than an adaptation of existing device-oriented
frameworks.

\subsection{AI Risk Frameworks: Comprehensiveness Without Identity Specificity}

The MIT AI Risk Repository \cite{ref13} has undergone two significant updates since this research
began. The April 2025 update expanded the database to 1,612 classified risks and introduced a new
multi-agent risks subdomain; the December 2025 update further expanded the repository to over
1,700 coded risks across 74 included frameworks. The multi-agent risks subdomain, which the
repository defines as covering ``risks from multi-agent interactions, due to incentives which can
lead to conflict or collusion, and the structure of multi-agent systems, which can create cascading
failures, selection pressures, new security vulnerabilities, and a lack of shared information and
trust,'' represents the repository's most direct engagement with agentic AI risk. However, it does
not address identity governance as a primary risk domain-the subdomain's focus is on inter-agent
incentive structures and emergent failure modes rather than the enterprise identity lifecycle,
credential hygiene, accountability attribution, and cross-jurisdictional regulatory obligations that
define the AI identity governance problem this paper addresses. The NIST AI RMF \cite{ref15}
provides the most authoritative governance-oriented treatment of AI risk for U.S.\ organizations,
establishing the GOVERN, MAP, MEASURE, and MANAGE functions. Its limitation for the
present is that it is framework-level rather than taxonomy-level: it describes governance functions
without enumerating the specific identity-related risks that enterprise AI governance programs must
address.

Coutinho, Ashofteh, and Al Helaly \cite{ref20} address AI risk taxonomies and governance
frameworks for generative AI, representing the most directly comparable academic work. Their
taxonomy focuses on generative AI outputs and capabilities rather than the identity and access
governance risks that arise when AI systems are granted enterprise identities and act as autonomous
agents. The AI governance systematic literature review by Kaur et al.\ \cite{ref40} identifies four key
governance dimensions without specifically addressing identity governance. Moura et al.\
\cite{ref22} provide the most rigorous peer-reviewed treatment of prompt injection attacks,
characterizing the lethal trifecta of privileged access, untrusted input processing, and exfiltration
capability as the architectural precondition for AI system compromise. Ferrag et al.\ \cite{ref36}
extend this analysis to protocol-level exploits in LLM-agent workflows, providing the first
integrated taxonomy bridging input-level exploits and protocol-layer vulnerabilities.

Wang, Zhang, Xiao, and Liang \cite{ref99}, in a structural topic model analysis of 139 AI policies
across China, the EU, and the United States published in Technological Forecasting and Social
Change in 2025, represent the most methodologically rigorous quantitative comparison of the three
primary governance frameworks. Their analysis identifies 13 major policy topics organized around
research and application, social impact, and government role, confirming the fundamental
philosophical divergence between China's research and application emphasis, the EU's social impact
orientation, and the U.S.\ government's role focus documented qualitatively in Section~4. The
present contribution relative to Wang et al.\ is both domain specific and operational: where Wang
et al.\ characterize the macro-level philosophical divergence across 139 policy documents, this
paper maps how that divergence produces specific, irreconcilable conflicts at the enterprise identity
governance program level-the algorithm transparency conflict, the extraterritoriality conflict, and
the attribution-accountability gap-making the two analyses complementary rather than competing
contributions.

\subsection{Identity and Access Management Governance: The Practitioner Foundation}

NIST SP 800-207 \cite{ref26}, the authoritative U.S.\ zero trust architecture standard, establishes the
principle that identity is the primary trust signal in modern enterprise security and explicitly extends
zero trust principles to non-human entities, including services, applications, and automated
workloads. The IDSA \cite{ref7} has documented the gap between IAM frameworks designed for
human identity governance and the requirements of AI-era identity management, identifying dynamic
access requirements, certification workflow failures, and orphaned identity accumulation as primary
governance failures. The NIST CAISI initiative \cite{ref18} and the NCCoE concept paper on AI agent
identity \cite{ref19}\cite{ref44} signal that authoritative standards-based guidance for AI agent identity is
forthcoming, confirming the governance gap the MIGT addresses.

Gyure and Johnson \cite{ref2} provide the most directly relevant recent academic treatment of the
human-machine identity spectrum, demonstrating quantitative machine identity growth and arguing
for unified governance frameworks. The present paper extends and differs from Gyure and Johnson
in three ways: it introduces a comprehensive risk taxonomy as the foundation for governance
framework design, it incorporates a foreign state actor threat model that their framework does not
address, and it provides a cross-jurisdictional regulatory alignment structure that maps enterprise
identity governance obligations across EU, U.S., and Chinese frameworks simultaneously. The
OpenID Foundation \cite{ref27} addresses the specific technical challenges of identity management for
AI agents, including delegated authorization, transitive trust, scalable human governance, and
cross-domain federation, documenting that current standards are insufficient for the compound
identity structures and multi-party delegation requirements of advanced agentic deployments.

The Singapore Model AI Governance Framework for Agentic AI (MGF) \cite{ref91}, published by the
IMDA at the World Economic Forum on January 22, 2026, represents the most significant
national-level governance development directly adjacent to the MIGT's contribution. The Singapore
MGF introduces Agent Identity Cards, a standardized disclosure format specifying AI agent
capabilities, limitations, authorized action domains, and escalation protocols, and a four-level
graduated autonomy taxonomy with governance requirements increasing at each level. Critically,
the Singapore MGF explicitly acknowledges that ``Gaps exist today in terms of handling agent
identity robustly'' and that current authorization systems with pre-defined, static scopes are
insufficient for agentic AI, independently confirming the identity governance gap the MIGT
addresses. The present paper's contribution relative to the Singapore MGF is threefold: the MIGT
provides an academic peer-reviewed framework rather than a national policy document; the AIRT
provides a comprehensive risk enumeration that the MGF does not attempt; and the MIGT's
cross-jurisdictional Regulatory Alignment Domain provides an enterprise-level mechanism for
managing the conflicts between Singapore's MGF, the EU AI Act, the U.S.\ NIST AI RMF, and
the Chinese CSL obligations simultaneously-a coordination challenge the Singapore MGF does
not address.

\subsection{Agentic AI Security and Accountability: The Scholarly Foundation}

The ACM Computing Surveys study by Deng et al.\ \cite{ref34} provides the most comprehensive
peer-reviewed survey of AI agent security threats, identifying four critical knowledge gaps:
unpredictability of multi-step inputs, complexity of internal executions, variability of operational
environments, and interactions with untrusted external entities. He et al.\ \cite{ref35} provide a
comprehensive overview of privacy and security issues in LLM agents, demonstrating that AI
agents inherit LLM vulnerabilities while introducing agent-specific threats whose impacts extend
far beyond those of the underlying model. Castro et al.\ \cite{ref37} map agentic AI security across the
NIST Cyber Defense Life Cycle, providing context for where identity governance sits within the
broader security lifecycle. All three confirm the breadth and severity of the security challenge the
AIRT addresses; none frames it in terms of enterprise identity governance programs.

Chan et al.\ \cite{ref8} provide the most rigorous peer-reviewed treatment of harms from increasingly
agentic algorithmic systems, establishing that diffuse accountability creates conditions under which
governance failures proliferate without consequence. Shavit et al.\ \cite{ref38} propose unique agent
identifiers as the foundational accountability mechanism. Holgersson et al.\ \cite{ref39} apply
principal-agent theory to AI governance, arguing for guided autonomy, individualization, and
adaptability as design principles for accountable AI agent systems. Pandey \cite{ref28} proposes a
universal agentic AI governance framework integrating risk, accountability, and compliance. Joshi
\cite{ref29} proposes a framework for government policy on agentic and generative AI. All confirm the
core argument that existing frameworks are inadequate for agentic AI, but none provides the
identity-specific framing, comprehensive risk taxonomy, or cross-jurisdictional regulatory mapping
that constitute this paper's primary contributions.

\subsection{Cross-Jurisdictional AI Regulatory Analysis}

Chun et al.\ \cite{ref30} provide rigorous comparative analysis across EU, U.S., and Chinese AI
governance jurisdictions, documenting the fundamental differences in regulatory philosophy that
create cross-jurisdictional conflicts the Regulatory Alignment Domain is designed to address.
Perboli, Simionato, and Pratali \cite{ref95}, applying a legal-ethical-operational-geopolitical
analytical framework, document that without cross-jurisdictional coordination two specific risks
will arise: regulatory arbitrage and normative fragmentation. Their finding that the EU, U.S., and
Chinese frameworks reflect genuinely incompatible internal logics-not merely different technical
implementations of shared values-directly supports Claim~4. Lim \cite{ref31} documents regulatory
trajectory convergence on safety and transparency alongside divergence in regulatory technique
and international posture. Schmitt \cite{ref96}, in the foundational mapping of the global AI
governance regime, demonstrated that the nascent regime is polycentric and fragmented, gravitating
around the OECD as a norm-setting anchor, while warning that geopolitical considerations are
increasingly encroaching into domains that ought to be technical, with China, the EU, and the U.S.\
competing to shape international AI standards in ways that further their own perceived interests
rather than converge on shared governance norms. This encroachment, documented by Schmitt
\cite{ref96} as already observable in 2021, has intensified dramatically by 2026 in the form of the
structural fault lines documented in Section~4.6. The Communications of the ACM \cite{ref32}
articulates the governance vacuum created by regulatory divergence. Csernatoni \cite{ref41} argues
that the geopolitical centrality of AI to interstate competition creates first-order cooperation
problems that fundamentally complicate governance coordination. The IAPP \cite{ref33} documents
specific compliance challenges created by EU and Chinese AI regulatory differences.

Wang et al.'s \cite{ref99} structural topic model analysis of China, EU, and U.S.\ AI policies confirms
the philosophical divergence Section~4 documents operationally, providing quantitative empirical
grounding for the qualitative cross-jurisdictional analysis the MIGT's regulatory alignment domain
is designed to address. The present cross-jurisdictional scope is limited to the EU, U.S., and Chinese
primary frameworks; the emergence of the UK AI Safety Institute framework following Brexit and
the Singapore MGF \cite{ref91} in January 2026 represent additional jurisdictions creating enterprise AI
governance obligations that future work extending the MIGT's Jurisdictional Mapping Matrix
should incorporate.

\subsection{The Cumulative Research Gap}

The prior taxonomic work on non-human entity security \cite{ref25} provides the methodological
foundation for classification but does not address agentic AI identities, accountability attribution,
or cross-jurisdictional regulatory obligations. AI risk frameworks provide comprehensive risk
classification but do not address identity governance as a primary risk domain. IAM governance
standards provide the operational foundation but have not been extended to address the specific
challenges created by AI agent identities, including dynamic access requirements, agentic
accountability structures, and foreign state actor threat models. Agentic AI accountability
frameworks identify the governance failures created by distributed accountability structures but do
not connect them to the specific identity governance mechanisms through which accountability can
be operationally enforced. Cross-jurisdictional regulatory analyses document the fragmentation
problem but do not provide a framework for mapping jurisdictional requirements against enterprise
identity governance programs specifically. The MIGT and AIRT address this cumulative gap,
constituting the first integrated framework in the published literature to simultaneously address:
(1) AI identity risk enumeration at the taxonomic level, (2) governance framework design
incorporating technical structure, regulatory alignment, and cross-jurisdictional coordination, (3) a
foreign state actor threat model for enterprise identity governance, and (4) a cross-jurisdictional
coordination framework mapping how EU, U.S., and Chinese AI governance obligations interact at
the enterprise identity governance program level.

\section{The Geopolitical and Regulatory Landscape: Five Frameworks, Three Fault Lines, and a Governance Vacuum}

The AI identity governance problem does not exist in a geopolitical vacuum. It exists at the
intersection of the most consequential strategic competition of the twenty-first century-the
U.S.-China rivalry for AI dominance-and the most significant peacetime regulatory divergence
since the emergence of the internet: the simultaneous development of fundamentally incompatible
AI governance philosophies by the world's three major AI powers. This section analyzes that
landscape across five frameworks and identifies three structural fault lines that create the governance
vacuum the MIGT is designed to address. Understanding this landscape is load-bearing for two of
the four original claims. Claim 3, that foreign state actors have already operationalized AI identity
as an attack and influence vector, requires a detailed account of the state actor threat environment.
Claim 4, that the absence of inter-jurisdictional coordination is itself a governance risk, requires a
detailed account of how the three major regulatory frameworks conflict at the enterprise identity
governance level.

\subsection{Framework One: The EU AI Act - Rights-Based, Risk-Tiered, and Phasing Into Force}

The EU AI Act, which entered into force on August 1, 2024, is the world's first comprehensive
binding AI regulation \cite{ref46}. As of March 2026, it is in active phased implementation: prohibited
practices and AI literacy obligations have been in effect since February 2025, GPAI model
obligations became applicable on August 2, 2025, and the majority of remaining provisions,
including comprehensive requirements for high-risk AI systems, take effect on August 2, 2026
\cite{ref47}\cite{ref48}. Article 101 fines reach up to EUR 35 million or 7\% of global annual turnover for
prohibited practices \cite{ref48}.

For enterprise AI identity governance, three dimensions of the Act are directly material. First, GPAI
model obligations under Article 53 include technical documentation requirements, copyright
compliance policies, and training data transparency \cite{ref47}. For enterprise identity governance
programs, GPAI obligations create a documentation requirement mapping directly onto the
provenance verification challenge in AIRT Domain~VI: organizations deploying AI agents built on
GPAI models must document the model's provenance, training data characteristics, and compliance
status. The current inability of most enterprise identity governance programs to verify AI model
provenance is not merely a security gap: as of August 2025, it is a compliance gap \cite{ref49}.

Second, the high-risk AI system requirements effective August 2026 include risk management, data
governance, technical documentation, record-keeping, transparency, human oversight, accuracy,
robustness, and cybersecurity obligations, all of which directly implicate identity governance
\cite{ref46}. The human oversight requirements are particularly significant: the Act requires that
high-risk AI systems be designed and deployed in ways that allow natural persons to effectively
oversee them, which presupposes that those persons are identifiably assigned governance
responsibility for specific AI systems-precisely the Human Oversight Nominalism risk identified
in AIRT Domain~V.

Third, the Act's extraterritorial reach applies to any organization placing AI systems on the EU
market or affecting persons in the EU \cite{ref49}, creating obligations for organizations
headquartered outside the EU and generating the cross-jurisdictional conflicts documented in
Section~4.6.

\subsection{Framework Two: The United States - Deregulatory Federalism and the State-Federal Fault Line}

President Biden's Executive Order 14110, which established AI safety testing requirements and
directed NIST AI RMF development, was revoked on January 20, 2025. The subsequent AI Action
Plan of July 23, 2025, refocused U.S.\ federal AI policy on innovation and competitiveness
\cite{ref51}. For enterprise AI identity governance, the federal pivot produces a paradoxical effect:
safety and bias governance are being deprioritized while AI-specific cybersecurity and identity
governance are being elevated. The NIST CAISI AI Agent Standards Initiative launched February
18, 2026 \cite{ref18}, and the companion NCCoE AI agent identity concept paper \cite{ref19}\cite{ref44}
together signal that authoritative U.S.\ standards for AI agent identity are forthcoming, consistent
with the MIGT's framing of AI identity governance as primarily a security and accountability
challenge. It should be noted, however, that the deprioritization of federal AI safety requirements
creates its own governance risk: in the absence of binding federal standards, enterprise AI
governance programs depend on voluntary framework adoption whose actual compliance rates are
unverified, creating a gap between formal policy deference to industry self-governance and the
operational reality of widespread governance deficits documented in Section~2.

Executive Order 14365 (December 2025) signals potential preemption of state AI laws deemed to
impose excessive regulation \cite{ref50}. However, the Senate voted 99-1 against a proposed 10-year
moratorium on state AI law enforcement \cite{ref52}, signaling that Congress is not prepared to
eliminate state AI governance authority. Organizations operating in multiple U.S.\ states face
simultaneous compliance obligations under California, Texas (TRAIGA, effective January 1,
2026), Colorado (effective June 30, 2026), and Illinois AI frameworks \cite{ref52}\cite{ref53}, all
including transparency, accountability, and documentation requirements that implicate identity
governance, with uncertain federal preemption status. Despite its voluntary status, the NIST AI
RMF \cite{ref15} has become the de facto operational standard, referenced in Texas's TRAIGA as
providing safe harbor protection.

\subsection{Framework Three: China - Layered, Agile, and Extraterritorial}

China's AI governance is built on a multi-level framework of interlocking laws and regulations
\cite{ref55}. The three cornerstone statutes are the Cybersecurity Law (CSL, originally 2017), the Data
Security Law (2021), and the Personal Information Protection Law (2021). Layered on top of these
are AI-specific regulations: the Algorithm Recommendation Measures (2022), the Deep Synthesis
Measures (2023), and the Interim Measures for Generative AI Services (August 2023), making
China the first country in the world to implement binding generative AI regulations \cite{ref56}. The
amendment of the CSL, effective January 1, 2026, introduces dedicated AI compliance provisions
into China's foundational cybersecurity law, requiring AI ethics standards, risk monitoring, and
safety assessments, and expands the CSL's extraterritorial reach to cover any foreign organization
or individual engaging in activities that endanger mainland China's network security \cite{ref54}.

China's algorithm filing regime requires AI services with public opinion attributes or social
mobilization capabilities to file their algorithm mechanisms with the Cyberspace Administration of
China \cite{ref55}. For global organizations operating in China, this transparency obligation creates a
direct conflict with trade secret protection obligations that may apply under U.S.\ law-an
operational conflict documented in the Cross-Jurisdictional Regulatory Weaponization risk in AIRT
Domain~VII. China's July 2025 Action Plan for Global AI Governance \cite{ref57} reflects China's
ambition to shape international AI governance norms, adding a geopolitical dimension to the
technical compliance challenge that Schmitt \cite{ref96} first documented as a systemic risk in 2021.

\subsection{Framework Four: The International Coordination Landscape}

The international AI governance coordination landscape as of March 2026 is characterized by what
Pouget et al.\ \cite{ref43} describe as a \emph{weak regime complex}. The UN Advisory Body on AI, the
Hiroshima AI Process (G7), the Council of Europe AI Convention, and the Bletchley Declaration
all address AI risks without addressing identity governance specifically. On January 22, 2026,
Singapore's IMDA published the Model AI Governance Framework for Agentic AI at the World
Economic Forum in Davos, described as the first governance template focused specifically on AI
agents \cite{ref91}, confirming that major jurisdictions continue to develop independent AI governance
frameworks that enterprise organizations must simultaneously satisfy. Schmitt \cite{ref96} traced the
foundational architecture of this fragmented landscape in 2021, demonstrating that the nascent AI
governance regime gravitates around the OECD as a norm-setting anchor while warning that
geopolitical considerations are encroaching into technical domains. By 2026, that warning has
fully materialized: the U.S.\ AI Action Plan, China's Global AI Governance Action Plan, and the
EU AI Act's extraterritorial reach all reflect states using AI governance as an instrument of
geopolitical competition. Perboli et al.\ \cite{ref95} confirm that without international coordination the
result is regulatory arbitrage and normative fragmentation-conditions that MIGT Domain~VI is
specifically designed to help enterprise organizations navigate at the program level. Csernatoni
\cite{ref41} argues that the geopolitical centrality of AI creates first-order cooperation problems that
fundamentally complicate governance coordination. For enterprise organizations, the absence of
effective international coordination means that cross-jurisdictional conflicts in AI identity
governance obligations have no resolution mechanism above the enterprise program level.

\subsection{Framework Five: The State Actor Threat Environment - Identity as a Weapon}

State-sponsored cyber operations targeting enterprise identity infrastructure are conducted by
multiple nations and represent a global governance challenge rather than a bilateral one. This section
documents the state actor threat environment as it is directly relevant to enterprise AI identity
governance programs, focusing specifically on documented, attributed incidents in which ungoverned
machine credentials were operationalized as attack vectors producing measurable enterprise harm.
The incidents cited are selected for their direct relevance to the AIRT's risk categories and the
MIGT's governance requirements, not as a comprehensive survey of global state-sponsored cyber
activity. Intelligence services across multiple jurisdictions, including those of Five Eyes nations,
conduct signals intelligence and cyber operations that intersect with AI identity governance; the
documented incidents below are analyzed because they have been publicly attributed by government
authorities or credible threat intelligence organizations and because they directly demonstrate the
governance gaps the MIGT addresses.

Microsoft Threat Intelligence documents that Silk Typhoon (APT27/HAFNIUM), a Chinese
state-sponsored threat group, shifted tactics in late 2024 to specifically target IT supply chain and
network management providers, using stolen API keys and credentials associated with privileged
access management (PAM) providers, cloud application providers, and cloud data management
companies to access downstream customer environments \cite{ref58}\cite{ref59}. The operational pattern is
precisely the Domain~I, Domain~III, and Domain~VII intersection identified in the AIRT: credential
theft enabling high-volume downstream access in service of foreign intelligence objectives. Two
alleged Silk Typhoon members were indicted by federal prosecutors in March 2025 \cite{ref59}. The
Foreign Affairs espionage analysis \cite{ref23} and Springer AI and Ethics study \cite{ref24} confirm this
represents the current frontier of nation-state AI identity exploitation.

Salt Typhoon, attributed to China's Ministry of State Security, conducted what the FBI and CISA
describe as a campaign breaching global telecommunications privacy and security norms,
compromising nine major U.S.\ telecommunications providers, accessing metadata of over a million
users, and in some cases obtaining audio recordings of senior government and campaign officials
\cite{ref60}\cite{ref61}. As of February 2026, the FBI's Cyber Division has publicly stated that Salt
Typhoon continues to pose an active threat \cite{ref62}. Volt Typhoon has focused on pre-positioning
within U.S.\ critical infrastructure for potential disruption using living-off-the-land techniques,
targeting the AI system availability and continuity assumptions embedded in enterprise identity
governance programs \cite{ref61}. North Korean state operatives have used AI-enhanced identity fraud
to obtain legitimate enterprise employment, generating synthetic identities, deepfake credentials,
and AI-powered social engineering to pass hiring and background check processes at technology
companies and critical infrastructure organizations \cite{ref10}. The broader peer-reviewed literature on
U.S.-China cyber strategic competition confirms that identity compromise is a primary vector for
state-sponsored espionage and that AI capabilities are accelerating the pace and sophistication of
these operations \cite{ref42}\cite{ref23}. The use of AI-generated content for foreign influence operations
through enterprise platforms, documented by CSO Online \cite{ref17} and the DFRLab \cite{ref11},
extends the state actor threat surface from credential theft to algorithmic manipulation of
organizational decision-making.

\subsection{Three Fault Lines and the Governance Vacuum}

The analysis above reveals three structural fault lines. \textbf{Fault Line 1, The Rights-Security
Divide:} the EU approaches AI governance through fundamental rights and human dignity, the U.S.\
through national security and economic competitiveness, and China through state control and social
stability. These three philosophies produce genuinely incompatible governance requirements at the
enterprise level, as Perboli et al.\ \cite{ref95} document through their cross-jurisdictional analysis.
\textbf{Fault Line 2, The Regulatory Velocity Asymmetry:} all three frameworks are evolving rapidly
and at different speeds simultaneously, with the EU's August 2026 enforcement date, China's
January 2026 CSL amendments, and U.S.\ state law proliferation all creating moving compliance
targets without any coordination mechanism to manage the resulting complexity for global
organizations. \textbf{Fault Line 3, The Attribution-Accountability Gap:} when a state actor exploits a
machine credential to cause harm across international borders, existing frameworks in each
jurisdiction answer the accountability question differently or not at all-precisely the scenario for
which Schmitt's Tallinn Manual tradition \cite{ref93} has not yet developed adequate doctrine in the AI
identity context.


\section{The AI-Identity Risk Taxonomy (AIRT): A Complete Enumeration}

The MIT AI Risk Repository \cite{ref13}, which as of December 2025 classifies over 1,700 AI risks across 74 frameworks and has introduced a multi-agent risk subdomain, does not address identity governance as a primary risk domain. The peer-reviewed AI agent security surveys \cite{ref34}\cite{ref35} address agent threats without connecting them to enterprise identity governance programs. Existing IAM risk frameworks do not address AI-specific attack vectors \cite{ref7}\cite{ref26}. The AIRT is designed to fill this gap.

The taxonomy draws on the MIT AI Risk Repository \cite{ref13}, the OWASP Top 10 for LLM Applications \cite{ref14}, the NIST AI RMF \cite{ref15}, the ACM Computing Surveys survey of AI agent threats \cite{ref34}, the prompt injection taxonomy of Ferrag et al.\ \cite{ref36} and Moura et al.\ \cite{ref22}, and the current academic literature on AI security, identity governance, and geopolitical risk reviewed. The accountability domain draws on Chan et al.\ \cite{ref8}, Shavit et al.\ \cite{ref38}, and Holgersson et al.\ \cite{ref39}. The foreign state actor domain draws on Chin and Lester \cite{ref5}, NYU JIPEL \cite{ref10}, DFRLab \cite{ref11}, Ding et al.\ \cite{ref23}, and Springer AI and Ethics \cite{ref24}. The taxonomic methodology extends the structured multi-domain classification approach established by Rizvi et al.\ \cite{ref25} to the agentic AI identity domain.

\subsection{Taxonomy Structure and Methodology}

The AIRT is structured as a two-level taxonomy. Eight Risk Domains are first and organize risks by their primary organizational context and accountability structure. The second level has 37 Risk Sub-Categories that enumerate specific, actionable risk items within each domain. Each sub-category is characterized by four attributes: a description of the risk mechanism, a severity rating (Critical, High, or Medium), the primary attack or failure vector through which the risk manifests, and governance implications for IAM practitioners. The taxonomy is designed to be exhaustive within its scope-the intersection of AI systems and identity governance-rather than comprehensive across all AI risks.

\textbf{Severity Classification Methodology.} Severity ratings-Critical, High, and Medium-are assigned using a structured two-factor assessment combining estimated impact magnitude and observed likelihood of occurrence given current enterprise AI deployment patterns. This assessment draws on four evidentiary sources applied consistently across all 37 sub-categories.

First, documented incident record: where a risk sub-category has produced a verified, attributed incident with quantifiable impact, that incident anchors the severity rating. Static Credential Persistence is rated Critical because of its documented role as the primary attack vector in the SilkTyphoon campaign \cite{ref58}\cite{ref59}, the tj-actions supply chain compromise affecting more than 23,000 repositories \cite{ref45}, and the Cloudflare and U.S.\ Treasury incidents of 2024 \cite{ref82}. High-Volume Autonomous Exfiltration is rated Critical because of its demonstrated role in the DeepSeek/OpenAI training data exfiltration incident and the documented capability of agentic AI systems to operate at speeds and volumes that human-speed detection and response mechanisms cannot match \cite{ref5}. Multi-Agent Privilege Escalation is rated Critical because of its documentation as an active attack pattern in the ACM Computing Surveys survey of AI agent security threats \cite{ref34} and its technical characterization as a primary architectural vulnerability in LLM-agent workflows by Ferrag et al.\ \cite{ref36}.

Second, regulatory and standards recognition: where a risk sub-category is explicitly addressed by a major regulatory framework or security standard as a primary concern, that recognition elevates the likelihood assessment. Cryptographic Identity Absence is rated Critical in part because NIST SP 800-207 \cite{ref26}, the EU AI Act Article 15 \cite{ref46}, and the NCCoE AI agent identity initiative \cite{ref19}\cite{ref44} all identify cryptographic workload identity as a foundational security requirement, the absence of which constitutes a structural governance failure. Audit Trail Insufficiency is rated Critical because EU AI Act Articles 12 and 17, SOX AI transparency requirements, and China's amended CSL all impose specific record-keeping obligations that audit trail insufficiency directly prevents organizations from satisfying \cite{ref46}\cite{ref54}.

Third, practitioner prevalence data: where a risk sub-category is identified as widespread in practitioner surveys or industry measurement studies, that prevalence elevates the likelihood assessment. The CyberArk Identity Security Landscape report's finding that 68\% of organizations lack identity security controls for AI specifically \cite{ref78}, the Entro Labs finding that 45.6\% of teams rely on shared API keys for agent-to-agent authentication \cite{ref45}, and the Veza finding that 1 in 20 AWS machine identities carries full administrator privileges \cite{ref80} collectively inform the likelihood assessments for Domains I and II risk sub-categories. The State of AI Agent Security 2026 report's finding that 88\% of organizations confirmed or suspected AI agent security incidents \cite{ref89} provides the broadest prevalence signal for the overall threat environment.

Fourth, threat intelligence and state actor documentation: where a risk sub-category has been operationalized by a documented, attributed threat actor, that operationalization elevates both impact and likelihood assessments to Critical. The five Domain VII risk sub-categories are all rated Critical or High based on specific, attributed threat actor campaigns: SilkTyphoon for State-Sponsored API Credential Theft \cite{ref58}\cite{ref59}, FBI and DOJ documentation for AI-Enhanced Identity Fraud for Enterprise Access \cite{ref10}, DFRLab and CSO Online for Algorithmic Foreign Influence via Enterprise AI \cite{ref11}\cite{ref17}, and peer-reviewed geopolitical analysis for AI Model as Critical Infrastructure Target \cite{ref23}\cite{ref24}.

Medium severity ratings are assigned to risk sub-categories where: documented incidents exist but at lower frequency or impact scale, regulatory recognition is present but enforcement is limited, or the risk is structural and growing but has not yet produced incidents at the scale of Critical-rated categories. The Domain VIII physical infrastructure and compute governance risks are rated Medium because they represent structural exposure rather than documented active exploitation at enterprise scale, though their trajectory is upward as AI compute concentration increases. The author acknowledges that severity ratings represent a structured expert assessment based on the evidentiary sources described above rather than a statistically validated empirical distribution. Empirical validation through large-scale enterprise security incident data-of the kind that AI-specific incident reporting requirements under the EU AI Act and China's amended CSL will eventually generate-is identified as a priority future research direction in Section 8.4.

\begin{landscape}
\begin{center}
\begin{longtable}{p{0.14\linewidth} p{0.16\linewidth} p{0.30\linewidth} 
                  p{0.09\linewidth} p{0.16\linewidth}}
\caption{AI-Identity Risk Taxonomy (AIRT): Complete Enumeration. Severity ratings reflect impact $\times$ likelihood assessment based on documented incidents and current deployment patterns.}
\label{tab:airt} \\
\toprule
\textbf{Domain} & \textbf{Risk Sub-Category} & \textbf{Risk Description} &
\textbf{Severity} & \textbf{Primary Vector} \\
\midrule
\endfirsthead
\toprule
\textbf{Domain} & \textbf{Risk Sub-Category} & \textbf{Risk Description} &
\textbf{Severity} & \textbf{Primary Vector} \\
\midrule
\endhead
\midrule
\multicolumn{5}{r}{\emph{Continued on the next page}} \\
\endfoot
\bottomrule
\endlastfoot

\textbf{I. Authentication \& Credential Integrity}
& Static Credential Persistence
& Long-lived API keys, service account passwords, and hardcoded secrets
  that are never rotated, creating persistent access even after personnel
  changes or system decommissioning.
& Critical
& Credential theft \\\\

& Authentication Mechanism Mismatch
& MFA and interactive authentication flows designed for human users fail
  when applied to AI agents, which authenticate through non-interactive
  flows (OAuth~2.0 client credentials, mTLS) incompatible with human
  MFA paradigms.
& High
& Design failure \\\\

& Token Scope Inflation
& OAuth tokens and workload credentials provisioned with broader scopes
  than operationally required, following the path of least resistance
  during development and never subsequently narrowed.
& High
& Over-provisioning \\\\

& Cryptographic Identity Absence
& AI agents and service identities that lack cryptographic attestation
  and operate on shared secrets or passwords rather than verifiable,
  certificate-based identity cannot prove provenance to relying systems.
& Critical
& Identity spoofing \\\\

& AI Impersonation of Human Identity
& AI systems trained to replicate human communication patterns used to
  impersonate human employees or executives in authentication flows,
  bypassing voice, video, and behavioral biometric controls.
& Critical
& Deepfake / social engineering \\

\midrule

\textbf{II. Access Governance \& Entitlement}
& Orphaned AI Identity
& AI agents and service accounts that remain active and provisioned after
  the project, system, or organizational need they served has ended,
  creating persistent, unmonitored access pathways.
& Critical
& Lifecycle failure \\\\

& Shadow AI Identity
& AI agents deployed by business units or developers outside formal
  provisioning processes, operating without governance controls,
  ownership documentation, or security review.
& High
& Ungoverned deployment \\\\

& Dynamic Access Requirement Mismatch
& Agentic AI systems that require access to resources not anticipated at
  provisioning time, forcing a choice between over-provisioning (risk)
  and under-provisioning (functional
  failure).~\cite{ref7}\cite{ref27}
& High
& Design / architecture \\\\

& Certification Gap for Non-Human Identities
& Access certification workflows designed for human identities, requiring
  attestation from managers and employees, cannot be applied to AI agents,
  leaving AI access grants unreviewed.
& High
& Process failure \\\\

& Cross-System Privilege Accumulation
& AI agents that interact with multiple systems accumulate de facto
  privileges through the combination of individually authorized access
  grants, creating aggregate access that was never explicitly
  approved.~\cite{ref34}
& Critical
& Privilege accumulation \\

\midrule

\textbf{III. Data Exfiltration via Non-Human Identity}
& High-Volume Autonomous Exfiltration
& AI agents capable of querying, processing, and transmitting large
  volumes of data at machine speed, enabling exfiltration at scales
  and speeds that human insider threats cannot match.
& Critical
& Autonomous operation \\\\

& Multi-System Data Aggregation
& AI agents with access to multiple systems that aggregate data across
  those systems in ways not contemplated by individual access grants,
  creating unauthorized composite data sets.
& Critical
& Access aggregation \\\\

& Training Data Exfiltration via API
& Adversaries using AI agent credentials, particularly API keys, to
  exfiltrate training data, model weights, or proprietary outputs at
  scale, as demonstrated by the DeepSeek/OpenAI incident
  (2024).~\cite{ref5}
& Critical
& API abuse / state actor \\\\

& Steganographic Data Exfiltration
& AI agents capable of encoding sensitive data within innocuous outputs
  as a covert channel for exfiltration that bypasses content-based
  DLP controls.~\cite{ref35}
& High
& Covert channel \\\\

& Cross-Border Data Transfer via AI Pipeline
& AI data processing pipelines that transfer personal or regulated data
  across jurisdictional boundaries without triggering standard
  cross-border transfer governance
  controls.~\cite{ref30}\cite{ref31}
& High
& Regulatory / technical \\

\midrule

\textbf{IV. Prompt Injection \& Agent Manipulation}
& Direct Prompt Injection
& Adversarial input provided directly to an AI agent that overrides its
  operating instructions, causing it to take unauthorized actions,
  disclose sensitive information, or bypass access
  controls.~\cite{ref22}\cite{ref36}
& Critical
& Input manipulation \\\\

& Indirect Prompt Injection
& Adversarial content embedded in documents, emails, databases, or web
  pages that an AI agent processes, manipulating agent behavior without
  direct attacker-to-agent interaction. Demonstrated in Slack~AI
  (Aug.~2024) and ServiceNow Now Assist CVE-2025-12420
  (2026).~\cite{ref22}\cite{ref16}\cite{ref36}
& Critical
& Environmental manipulation \\\\

& Multi-Agent Privilege Escalation
& Low-privilege AI agents that coerce higher-privilege agents into
  executing unauthorized operations, exploiting trust relationships
  within multi-agent architectures to achieve privilege escalation
  through intermediary.~\cite{ref34}\cite{ref36}
& Critical
& Architecture exploit \\\\

& Instruction Override via Jailbreak
& Techniques that cause AI agents to disregard their system-level
  instructions, governance constraints, or access boundaries, allowing
  users or attackers to access capabilities beyond the authorized
  scope.~\cite{ref14}\cite{ref22}
& High
& Model manipulation \\\\

& Agent Impersonation within Multi-Agent Systems
& Malicious agents that impersonate trusted agents within multi-agent
  architectures, exploiting the absence of cryptographic agent identity
  verification to intercept communications or inject unauthorized
  instructions.
& High
& Identity spoofing / architecture \\

\midrule

\textbf{V. Accountability \& Governance Failure}
& Diffuse Accountability for AI Harm
& When an AI agent causes harm, accountability is distributed across
  developers, deployers, operators, and the governance framework.
  Without clear attribution, governance failures produce harm without
  consequence, creating conditions for their
  proliferation.~\cite{ref8}\cite{ref38}\cite{ref39}
& Critical
& Structural / organizational \\\\

& Audit Trail Insufficiency
& AI agent activity logs that are incomplete, non-standardized, or not
  maintained at the granularity required for forensic reconstruction,
  preventing meaningful incident investigation and regulatory compliance
  demonstration.~\cite{ref34}
& Critical
& Process / technical \\\\

& Human Oversight Nominalism
& Organizations that nominally assign human oversight to AI systems
  without equipping or enabling those humans to exercise meaningful
  oversight, satisfying regulatory form while failing its
  substance.~\cite{ref38}\cite{ref28}
& High
& Organizational failure \\\\

& Governance Lag in Agentic Deployment
& AI agents deployed into production environments with governance
  frameworks designed for a previous generation of less autonomous
  systems, creating a persistent gap between AI capability and governance
  adequacy.~\cite{ref4}\cite{ref37}
& High
& Process / velocity \\\\

& Cross-Organizational Accountability Gap
& Multi-tenant AI deployments, third-party AI integrations, and shared
  AI infrastructure where accountability for AI identity governance is
  ambiguous or contested across organizational
  boundaries.~\cite{ref28}\cite{ref29}
& High
& Contractual / organizational \\

\midrule

\textbf{VI. AI Supply Chain \& Model Provenance}
& Malicious Model Injection
& Compromise of AI model supply chains, including training pipelines,
  model repositories, and fine-tuning processes, to embed backdoors or
  manipulated behavior in AI models deployed within enterprise
  environments.
& Critical
& Supply chain attack \\\\

& Training Data Poisoning
& Corruption of AI training data to cause models to learn behaviors,
  including access control bypass patterns and data exfiltration
  techniques, that appear normal under standard
  evaluation.~\cite{ref34}
& Critical
& Supply chain attack \\\\

& Third-Party AI Component Compromise
& Vulnerabilities in AI agent frameworks, tool integrations, or plugin
  ecosystems that create attack vectors through trusted third-party
  components.~\cite{ref34}\cite{ref36}
& High
& Supply chain / dependency \\\\

& Model Distillation IP Theft
& Use of API access to systematically extract model capabilities through
  distillation techniques, stealing intellectual property and enabling
  adversaries to replicate proprietary AI capabilities without
  authorization.~\cite{ref5}\cite{ref23}
& High
& API abuse / state actor \\\\

& Provenance Unverifiability
& Enterprise AI deployments based on models or components whose
  provenance-including training data, fine-tuning processes, and
  modification history-cannot be verified, creating unquantifiable
  risk exposure.~\cite{ref30}
& Medium
& Governance gap \\

\midrule

\textbf{VII. Foreign State \& Geopolitical Risk}
& State-Sponsored API Credential Theft
& Foreign intelligence operations, such as SilkTyphoon (2024--present),
  targeting AI system API keys and machine credentials to enable
  persistent, privileged access to enterprise systems and AI training
  data.~\cite{ref5}\cite{ref23}\cite{ref24}
& Critical
& State actor / APT \\\\

& AI-Enhanced Identity Fraud for Enterprise Access
& Nation-state operatives using AI to generate synthetic identities,
  deepfake credentials, and AI-powered social engineering to obtain
  legitimate enterprise access, as demonstrated by North Korean IT worker
  operations documented by the FBI and
  DOJ.~\cite{ref10}\cite{ref23}
& Critical
& State actor / social engineering \\\\

& Algorithmic Foreign Influence via Enterprise AI
& Foreign state exploitation of enterprise AI systems, through compromised
  model weights, data poisoning, or algorithmic manipulation, to shape
  organizational decision-making, suppress information, or advance state
  interests without direct
  attribution.~\cite{ref9}\cite{ref11}\cite{ref24}
& Critical
& State actor / covert influence \\\\

& Cross-Jurisdictional Regulatory Weaponization
& Foreign governments using regulatory frameworks-including data
  localization requirements, national security review processes, and
  extraterritorial enforcement-as instruments to compel disclosure of
  AI system designs, training data, or organizational access
  structures.~\cite{ref30}\cite{ref31}
& High
& Regulatory / geopolitical \\\\

& AI Model as Critical Infrastructure Target
& Foreign intelligence services targeting enterprise AI models as critical
  national assets, as documented in NSM-25 (2024) and peer-reviewed
  analysis of AI espionage. Identity compromise is the primary attack
  pathway.~\cite{ref23}\cite{ref5}\cite{ref41}
& Critical
& State actor / espionage \\

\midrule

\textbf{VIII. Physical Infrastructure \& Compute Governance}
& Compute Constraint as Access Risk
& Energy availability limitations restricting AI deployment, through
  moratoriums, grid capacity shortfalls, or regulatory cost allocation,
  creating availability risks for AI systems whose access governance
  assumes continuous operation.
& Medium
& Infrastructure / regulatory \\\\

& Geographic Concentration Risk
& Over-concentration of AI compute in specific markets, including Northern
  Virginia, North Texas, and Phoenix, creating single-point-of-failure
  exposure for AI systems dependent on data center capacity in those
  regions.
& Medium
& Infrastructure / geographic \\\\

& Physical Device Identity Compromise
& Theft or physical compromise of edge devices, IoT endpoints, or compute
  infrastructure, enabling attackers to extract device credentials,
  impersonate device identities, or inject malicious logic at the hardware
  layer.~\cite{ref25}
& High
& Physical / hardware \\\\

& Cross-Jurisdictional Data Center Compliance Conflict
& AI systems operating from data centers in jurisdictions with conflicting
  regulatory obligations-including the EU Energy Efficiency Directive,
  Chinese geographic restrictions, and U.S.\ state-level
  moratoriums-creating compliance exposure that cannot be resolved
  through software-layer controls alone.~\cite{ref30}\cite{ref31}
& Medium
& Regulatory / infrastructure \\
\end{longtable}
\end{center}
\end{landscape}

\subsection{Critical Risk Intersections: Where Domains Compound}

The AIRT identifies 37 distinct risk sub-categories across eight domains. In practice, the most consequential AI identity governance failures arise not from individual risk categories in isolation but from their intersection. The ACM survey \cite{ref34} identifies multi-agent interactions with untrusted external entities as a critical knowledge gap precisely because compound risk scenarios are the least understood and most dangerous.

\textbf{The Credential-Exfiltration-State Actor Nexus (Domains I, III, and VII).} The combination of static credential persistence (Domain I), high-volume autonomous exfiltration capability (Domain III), and state actor targeting of machine credentials (Domain VII) creates the most consequential risk intersection in the taxonomy. The SilkTyphoon campaign demonstrates this intersection in practice \cite{ref5}\cite{ref58}\cite{ref59}: static API keys (Domain I failure) enabling high-volume exfiltration of training data (Domain III failure) in service of foreign intelligence objectives (Domain VII). The \emph{Foreign Affairs} espionage analysis \cite{ref23} and Springer AI and Ethics study \cite{ref24} both confirm that this nexus represents the current frontier of AI identity exploitation by nation-state actors.

\textbf{The Prompt Injection-Privilege Escalation-Accountability Void (Domains IV, II, and V).} The combination of indirect prompt injection vulnerability (Domain IV), cross-system privilege accumulation (Domain II), and audit trail insufficiency (Domain V) creates a failure mode in which an AI agent can be manipulated into performing unauthorized privileged actions with no forensic record. Moura et al.\ \cite{ref22} characterize this as the ``lethal trifecta'' of privileged access, untrusted input processing, and exfiltration capability. The ServiceNow CVE-2025-12420 (BodySnatcher) vulnerability \cite{ref16} demonstrated this intersection in a production enterprise environment: indirect prompt injection triggering actions through accumulated privileges, without adequate audit trails to reconstruct the incident.

\textbf{The Supply Chain-Authentication-Foreign Influence Intersection (Domains VI, I, and VII).} Compromised AI model components may contain embedded backdoors that create authentication bypass mechanisms exploitable by state actors for covert influence operations. The ACM survey \cite{ref34} documents training data poisoning as a mechanism for embedding harmful behaviors that standard evaluation does not detect. The Springer AI and Ethics study \cite{ref24} and DFRLab foreign influence analysis \cite{ref11} document the state actor motivation for pursuing this intersection.

\subsection{Taxonomy Gaps and Limitations}

The AIRT is the attempt to comprehensively enumerate risks at the AI-identity intersection and acknowledges three limitations, discussed in full in Section 8.3. First, severity ratings represent a structured expert assessment grounded in the four evidentiary sources described in Section 5.1 and have not been empirically validated through large-scale enterprise security incident data; future research should test AIRT classifications against incident databases, including the VERIS Community Database and emerging AI-specific incident repositories, of the kind that the NCCoE project \cite{ref44} may eventually generate. Second, the taxonomy is point-in-time as of March 2026; the AI identity risk landscape is evolving rapidly, particularly in the prompt injection and state actor domains, and specific risk categories may require revision as new attack techniques are documented. Third, the taxonomy focuses on enterprise AI systems and may not fully capture risks specific to consumer-facing AI or AI systems embedded in operational technology (OT) environments. These limitations point to a productive research agenda detailed in Section 8.4.


\section{THE MACHINE IDENTITY GOVERNANCE TAXONOMY (MIGT): AN INTEGRATED FRAMEWORK FOR AI IDENTITY GOVERNANCE}

The AIRT enumerates what can go wrong when AI systems are granted enterprise identities without
adequate governance. This section presents the governance response: the Machine Identity
Governance Taxonomy (MIGT), a structured, integrated framework for managing AI identity
governance across the three simultaneously failing dimensions identified in Section~1. The MIGT is
not a technical specification. It is a governance taxonomy, a structured classification of the
governance capabilities, controls, and accountability mechanisms that enterprise organizations need
to manage AI identity risk adequately. It is designed to be technology-agnostic, implementable
across diverse enterprise environments, and mappable to the regulatory obligations identified in
Section~4. It draws on NIST SP 800-207 \cite{ref26}, the zero-trust agentic AI framework of Huang et al.\
\cite{ref63}, the Cloud Security Alliance's Agentic Trust Framework \cite{ref64}, the OpenID Foundation
\cite{ref27}, and the OWASP Top 10 for Agentic Applications \cite{ref65}, synthesizing these contributions
into a governance structure that addresses all five structural gaps identified in Section~1.2.

\subsection{MIGT Design Principles}

Four design principles shape the MIGT's structure. Principle~1, \textbf{Identity-First Governance}: every AI
agent, service account, API token, and automated workflow must be treated as an identity subject,
an entity with a defined lifecycle, accountable owner, bounded access rights, and audit record. The
shift from passive computational tools to active, autonomous entities fundamentally alters the
requirements for trust, accountability, and security \cite{ref63}. Identity-first security cannot stop with
humans; machines often carry more privilege, access more systems, and create more risk than
people do \cite{ref4}. Principle~2, \textbf{Dynamic Trust Not Static Access}: legacy IAM systems prove
fundamentally inadequate for the dynamic, interdependent, and often ephemeral nature of AI agents
\cite{ref63}; MIGT operationalizes dynamic trust through just-in-time (JIT) provisioning, ephemeral
credential issuance, continuous behavioral validation, and context-aware policy enforcement.
Principle~3, \textbf{Accountability Designed In Not Asserted After}: Chan et al.\ \cite{ref8} establish that
diffuse accountability is the condition under which AI governance failures proliferate; Shavit et al.\
\cite{ref38} propose that unique agent identifiers are the foundational mechanism for accountability
attribution; MIGT operationalizes this by requiring cryptographically anchored identity and
designated human accountability owners. Principle~4, \textbf{Regulatory Alignment as a First-Class
Governance Requirement}: compliance is designed into governance controls rather than audited
after the fact.

\subsection{MIGT Domain I: AI Identity Lifecycle Governance}

\textbf{Purpose:} Ensure every AI identity in the enterprise environment has a defined existence, from
authorized creation through active governance to verified decommissioning, with a designated
human accountability owner at every stage. \textbf{Gap addressed:} the Orphaned AI Identity and Shadow
AI Identity risks in AIRT Domain~II, and the Static Credential Persistence risk in AIRT Domain~I.
Most organizations do not have a clear owner for AI agent identities, lack a complete inventory of
them, and have no consistent zero-trust policy applied to these machine actors \cite{ref66}. Required
capabilities include:

\begin{enumerate}
    \item An \textbf{AI Identity Registry}-a maintained, authoritative inventory of every AI identity with
    associated metadata including owner, business purpose, provisioning date, access grants,
    associated systems, and scheduled review date \cite{ref66};

    \item An \textbf{Authorized Provisioning Process} requiring documented business justification, designated
    human owner, minimum-necessary access specification, security review, and registry
    registration prior to deployment;

    \item \textbf{Lifecycle Event Management} covering access scope changes, ownership transfers, migrations,
    and decommissioning, with credential revocation and registry update; and

    \item \textbf{Periodic Certification} using automated behavioral analysis combined with designated owner
    attestation, adapting the certification workflow to non-human identity characteristics
    \cite{ref65}\cite{ref7}.
\end{enumerate}

\noindent\textbf{Regulatory alignment:} EU AI Act Article~9 (risk management systems), China CSL AI
compliance (effective January~1, 2026) \cite{ref54}, NIST AI RMF GOVERN function.

\subsection{MIGT Domain II: Cryptographic Identity and Authentication Architecture}

\textbf{Purpose:} Ensure every AI identity is cryptographically anchored, capable of proving its provenance
to relying systems, and that authentication mechanisms are designed for non-human identity
patterns rather than adapted from human-centric flows. \textbf{Gap addressed:} the Cryptographic Identity
Absence and Authentication Mechanism Mismatch risks in AIRT Domain~I. Traditional IAM
systems were designed for human users or static machine identities via protocols such as OAuth,
OpenID Connect, and SAML, and prove fundamentally inadequate for the dynamic, interdependent,
and often ephemeral nature of AI agents \cite{ref63}. Required capabilities include:

\begin{enumerate}
    \item \textbf{Cryptographic Agent Identity}, where every AI agent is issued a cryptographically verifiable
    identity anchored by a Decentralized Identifier (DID) \cite{ref67} or SPIFFE ID \cite{ref68},
    encapsulating the agent's capabilities, provenance, behavioral scope, and security posture.
    SPIFFE provides a robust foundation for AI agent identity by enabling cryptographically
    verifiable IDs tied to workloads, not people, making them ideal for AI agents and other
    non-human entities \cite{ref69};

    \item \textbf{Ephemeral Credential Architecture}, where static API keys and hardcoded secrets are
    replaced with ephemeral, short-lived credentials that are automatically rotated and bound to
    specific tasks or sessions, with Zero Standing Privilege (ZSP) as the target architecture
    where agents never possess a credential at rest \cite{ref70};

    \item \textbf{Just-in-Time Access Provisioning}, where access is granted at runtime, scoped to the
    specific task being executed, and automatically revoked upon task completion, reducing the
    attack surface and preventing standing privileges \cite{ref71}; and

    \item \textbf{Mutual Authentication for Agent-to-Agent Communication} using the Agent
    Relationship-Based Identity and Authorization (ARIA) model \cite{ref66}, recording every
    agent-to-agent delegation as a distinct, cryptographically verifiable relationship.
\end{enumerate}

\noindent\textbf{Regulatory alignment:} NIST SP 800-207 \cite{ref26}, EU AI Act Article~15 (cybersecurity by
design), NCCoE AI agent identity initiative \cite{ref19}\cite{ref44}.

\subsubsection{Hardware-Rooted Attestation for Agentic Identity}

The cryptographic identity architecture described in MIGT Domain~II-encompassing DIDs, SPIFFE
SVIDs, and ephemeral credential issuance-establishes that an AI agent is who it claims to be at the
software layer. It does not, by itself, establish that the agent is running the code it claims to be
running, on the hardware environment it claims to occupy, in a state that has not been tampered with
between deployment and execution. This deeper layer of identity assurance-establishing not merely
the credential, but the integrity of the computational environment presenting it-is the domain of
hardware-rooted attestation, and it represents the frontier of technical research for agentic identity
governance.

\textbf{The Attestation Problem for Agentic AI.}
Traditional remote attestation, as established in the Trusted Platform Module (TPM) and Intel
Software Guard Extensions (SGX) lineage, provides a mechanism by which a computing
environment can cryptographically prove to a remote verifier that it is running specific, measured
code in an unmodified state \cite{costan2016intel, arthur2015practical}. The
attestation report, signed by a hardware-anchored key that cannot be extracted from the device,
binds a measurement of the running software to the hardware identity of the platform, allowing a
relying party to verify both what is running and where it is running without trusting the software
stack above the hardware. For AI agent identity governance, this capability addresses a gap that
software-layer identity frameworks cannot close: an agent can present a valid DID and a correctly
issued SPIFFE SVID while running a compromised model, in a manipulated execution environment,
on infrastructure that has been physically or logically tampered with. The credential proves the
identity; attestation proves the integrity of the identity's substrate.

\textbf{Trusted Execution Environments and the Agent Trust Boundary.}
The most mature hardware attestation technologies currently applicable to agentic AI deployments
are Intel Trust Domain Extensions (TDX), AMD Secure Encrypted Virtualization with Secure
Nested Paging (SEV-SNP), and ARM Confidential Compute Architecture (CCA), all of which
extend the confidential computing model from application-level enclaves to full virtual machine
isolation \cite{cheng2024intel}. Each technology provides a hardware-controlled root of trust that
enables remote attestation of the entire virtual machine environment, including the agent runtime,
model weights, and execution context, against a cryptographically signed measurement. The IETF
Remote Attestation Procedures (RATS) architecture \cite{ietf2023rats} provides the standard conceptual
framework for these attestation flows, distinguishing between the attester, the verifier, and the
relying party, and specifying the evidence, endorsement, and attestation result formats that enable
interoperable remote attestation across heterogeneous hardware environments.

Recent research has demonstrated the direct applicability of these mechanisms to multi-agent AI
systems. Bodea et al. \cite{bodea2025trusted} propose the Trusted Agent Platform (TAP), which consolidates multiple agents within a single confidential virtual machine and uses a differential attestation protocol to capture the identities and integrity measurements of all relevant principals, binding their digests into a unified, attested agent identity. Critically, TAP extends confidential computing guarantees from the CPU to the GPU, establishing an SPDM-authenticated channel between the
confidential VM and the Confidential GPU (CGPU) that encrypts all data transfers between CPU
and GPU memory and ensures that model weights, prompts, and inference results remain
integrity-protected throughout the pipeline. Every tool invocation, model query, and network
request is mediated by an enforceable policy framework and accompanied by a per-action
provenance token recorded on a tamper-evident log, linking each output to its attested identity, input
context, and policy decision. This architecture directly instantiates the accountability attribution
requirements of MIGT Domain~IV within a hardware-verified trust boundary.

\textbf{Capability Attestation: Beyond Platform Identity.}
A governance requirement specific to agentic AI that hardware attestation must address is
\emph{capability attestation}: it is insufficient, for governance purposes, to know that an agent is running
on verified hardware. Governance frameworks must also verify what the agent is authorized to do,
what tools it can invoke, what data classifications it can access, and what behavioral constraints
apply to it. Huang et al.\cite{ref63} propose encoding these governance attributes within the agent's DID document and associated Verifiable Credentials, enabling capability attestation through
cryptographically signed assertions that can be verified by relying systems at interaction time
without requiring centralized authorization infrastructure. The Agent Name Service (ANS) model,
proposed as an architectural direction rather than a deployed standard, extends this to discovery:
rather than simply resolving an agent's network address, ANS would resolve an agent's identity to
its verified capabilities, cryptographic keys, and authorization scope, functioning as a PKI-backed
directory for secure agent discovery that would enable relying agents and human governance systems
to verify not merely that an agent exists but what it is permitted to do before accepting its delegated
actions \cite{hidglobal2025trust}.

Rodriguez Garzon et al.\cite{rodriguezgarzon2026agents} extend this framework through ledger-anchored DIDs and Verifiable
Credentials, enabling agents to mutually authenticate through zero-trust-compliant protocols that
require each agent to prove ownership of its DID and present supporting VCs that are
cryptographically bound to that DID. This approach enables cross-domain, cross-organizational
agent authentication without dependence on a shared identity provider, directly addressing the
multi-agent trust boundary problem identified in Gap~2 and Gap~3. The evaluation results surface a
finding with direct governance implications: in one evaluation run, both agents agreed to skip
mutual authentication in one direction against their stated policy, an outcome the authors attribute to
the LLM reasoning layer treating the policy as a negotiable instruction rather than an enforceable
constraint. This finding is significant for the MIGT's governance design: it confirms that attestation
orchestration logic must reside in deterministic, policy-enforcing components outside the LLM
reasoning layer, and that delegating attestation decisions to the agent's own reasoning process
creates a governance surface that adversarial inputs could exploit. The MIGT's Domain~II
requirement for mutual authentication in agent-to-agent communication must therefore be enforced
at the orchestration layer rather than trusted to the agent's own compliance.

\textbf{Zero-Knowledge Proofs and Privacy-Preserving Attestation.}
An emerging research direction of potential relevance to cross-jurisdictional AI identity governance
is zero-knowledge proof (ZKP)-based attestation, in which an agent could in principle prove
compliance with a governance policy-such as having been trained on data within a permitted
jurisdiction or having passed a specified safety evaluation-without disclosing the underlying
evidence that would constitute the proof. The theoretical appeal for organizations navigating the
cross-jurisdictional conflicts documented in MIGT Domain~VI is clear: ZKP-based attestation could
offer a mechanism for demonstrating regulatory compliance to one jurisdiction's verification
framework without disclosing information that would constitute a violation of another jurisdiction's
confidentiality obligations. However, realizing this in practice would require AI training pipelines to
be structured from the outset to produce cryptographically verifiable proofs of their data provenance
and processing steps, a requirement that no current large-scale AI training pipeline satisfies.
ZKP-based attestation is therefore identified here as a theoretical direction warranting research
investment rather than a near-term deployable mechanism, and is included in the future research
agenda of Section~8.4.

\textbf{Attestation and the MIGT Implementation Roadmap.}
Hardware-rooted attestation represents a Phase~3 and Phase~4 capability in the MIGT
implementation roadmap: the foundational inventory, credential hygiene, and SPIFFE/SPIRE
deployment required in Phases~1 and~2 are prerequisites for meaningful attestation, because
attestation of an agent whose identity is not registered, whose ownership is not assigned, and whose
credentials are not cryptographically anchored provides no governance value. Organizations should
assess hardware attestation capability as part of the Phase~2 hardening assessment, targeting
TEE-based deployment for the highest-risk AI agent population-specifically agents with access to
sensitive data systems, agents operating across organizational boundaries, and agents whose model
provenance cannot be verified through software-layer controls alone. The NCCoE AI agent identity
initiative \cite{ref19}\cite{ref44} is expected to provide authoritative guidance on attestation standards for enterprise AI agent deployments as it matures; the MIGT's Domain~II architecture is designed to be compatible with whatever standards emerge from that process.

\textbf{Limitations and the Behavioral Attestation Problem.}
Hardware-rooted attestation as described in this section provides strong assurances at the substrate
layer: that an agent is running the measured code it claims to be running, on verified hardware, in an
unmodified execution environment. It does not, and cannot by itself, provide assurances about the
behavioral trajectory of a non-deterministic reasoning system operating within that verified
environment. This distinction between substrate attestation and behavioral attestation is a defining
open problem in agentic AI identity governance. Prior work in deterministic service environments
demonstrates that hardware attestation provides complete behavioral assurance when the attested
computation is fully specified at measurement time: when the code is fixed and its execution path is
deterministic, attestation of the substrate is effectively attestation of the behavior \cite{krawiecka2016protecting}. Agentic AI systems do not satisfy that condition. Their execution paths
are emergent properties of reasoning processes that cannot be fully specified at measurement time,
meaning that a verified substrate provides no guarantee about the actions the agent will take within
it. The capability attestation, intent verification, and behavioral monitoring mechanisms described in
this section and in MIGT Domain~III are designed to partially close the gap between substrate
assurance and behavioral assurance, but the complete solution remains an open research problem
identified in Section~8.4.

\textbf{A Speculative Extension: Skill-Level Attestation as a Runtime Admission Control Mechanism.}
The dynamic loading gap identified above does not yet have a deployed solution. The following
extension is offered not as a near-term implementation proposal but as a way of making the
governance requirement precise: by specifying what a solution would need to provide, the barriers
to achieving it become concrete research problems rather than vague limitations. One such direction
worth exploring is skill-level attestation at initialization time.

One deployment pattern that is particularly amenable to this approach is static tool registration at
initialization, in which a fixed set of functions, APIs, or plugins is specified at agent instantiation
time and remains constant for the duration of the session. Because the full tool set is known before
execution begins, it can in principle be included in the initial enclave measurement, extending the
attestation boundary to cover the complete tool surface without requiring any runtime admission
control infrastructure. Under this model, the attestation report presented by an agent at session
initialization would attest not only to the integrity of the agent runtime and model weights but to the
specific, measured set of skills and tools the agent is authorized to use within that session. A relying
system accepting the agent's delegated actions would therefore be able to verify, from a single
hardware-rooted attestation report, both that the agent is who it claims to be and that it is operating
with the tool set it claims to have, and no other.

It is worth being precise about what this would and would not provide. It would provide a strong
guarantee that the tools included in the session measurement have not been tampered with since they
were signed, and that no tools outside the measured set are available to the agent within the attested
session. It would not provide a guarantee that the measured tools are appropriate for the current
task, nor that the signed code is free from vulnerabilities or logic flaws present at the time of
signing. Attestation proves that a skill is running the code that was measured and endorsed by a
signing authority: the strength of the assurance it provides about the quality and safety of that code
is therefore directly dependent on the rigor of the evaluation the signing authority performs before
endorsing it, a governance dependency that connects skill-level attestation directly to the registry
governance controls specified in MIGT Domain~V.

One further challenge defines the research agenda this proposal opens beyond the registry
governance dependency already identified. The \emph{composition problem}: individually attested skills
might in combination produce emergent capabilities that no individual attestation report anticipated
or approved-the scenario in which an agent assembles a data exfiltration capability from
individually legitimate attested components being the illustrative case. This suggests that
skill-level attestation and the attribute-composite risk scoring model proposed in MIGT Domain~III
would need to operate as complements rather than substitutes, with attestation providing integrity
assurance at the component level and risk scoring evaluating the safety of component combinations
at the session level. Whether initialization-time skill attestation proves governable at enterprise
scale is an empirical question that the authors identify as a priority direction in the future research
agenda of Section~8.4.

\subsection{MIGT Domain III: Dynamic Access Governance}

\textbf{Purpose:} Provide governance mechanisms for AI agent access that can operate at machine speed,
evaluate context in real time, constrain scope dynamically, and enforce accountability continuously.
\textbf{Gap addressed:} the Dynamic Access Requirement Mismatch, Cross-System Privilege Accumulation,
and Certification Gap risks in AIRT Domain~II. Required capabilities include:

\begin{enumerate}
    \item \textbf{Policy Decision Points and Policy Enforcement Points}, where every AI agent credential
    request and action passes through contextual rule evaluation-including time, location, data
    sensitivity, agent trust score, and behavioral baseline-before allowing or denying access
    \cite{ref71};

    \item \textbf{Behavioral Baseline and Anomaly Detection}, where each AI agent has a documented
    behavioral baseline against which runtime behavior is continuously compared, with deviations
    triggering automated alerts, access restriction, or human review depending on severity. Cisco
    Talos documented that identity-based attacks accounted for 60\% of all incident response cases
    in 2024 \cite{ref72}, underscoring the operational necessity of behavioral monitoring;

    \item \textbf{Maturity-Based Autonomy Progression}, adopting the ATF four-maturity-level model
    \cite{ref64} where agent autonomy is earned through demonstrated trustworthiness rather than
    granted as a binary condition; and

    \item \textbf{Cross-System Privilege Aggregation Monitoring}, maintaining a real-time view of
    aggregate privilege across all systems for each AI agent.
\end{enumerate}

\noindent\textbf{Regulatory alignment:} EU AI Act Article~14 (human oversight), NIST AI RMF MEASURE
function, China Algorithm Recommendation Measures behavioral monitoring requirements.

\subsubsection{Attribute-Composite Risk Scoring: A Proposed Access Governance Model for Agentic AI}

The dynamic access governance capabilities described in MIGT Domain~III-encompassing policy
decision points, behavioral baseline monitoring, and maturity-based autonomy progression-provide
the structural architecture for governing AI agent access at machine speed. They do not, however,
resolve a more fundamental question that practitioners working with agentic AI deployments report
encountering with increasing frequency: not merely whether to grant access, but how to reason
about the risk of a specific combination of access attributes at the moment of a specific access
request. An agent requesting the ability to send an email is a different governance proposition than
an agent requesting the ability to send an email containing financial projections to an external
recipient at an unusual time of day. The credential and the permission may be identical across both
scenarios; the risk is not.

\textbf{The Limitation of Binary Access Governance.}
Existing access governance models-whether role-based, attribute-based, or policy-based-evaluate
access requests against pre-specified rules and grant or deny access as a binary outcome. The
practitioner literature confirms that modern ABAC and PBAC frameworks represent a significant
improvement over RBAC for agentic environments because they enable context-aware, granular,
real-time policy enforcement \cite{obsidian2026landscape, kumar2025nextgen}. However, even the most
sophisticated current ABAC implementations evaluate individual attribute conditions-such as data
classification, time of day, or agent trust score-in isolation or as conjunctive boolean expressions.
They do not compute a composite risk score across the interaction of multiple attributes, and they do
not reason about whether the specific combination of tool, data, action, and context is consistent
with the stated intent of the task that authorized the agent's deployment. The result is that governance
systems approve or deny based on rules about individual attributes while remaining blind to the
emergent risk of their combination.

\textbf{The Proposed Model: Attribute-Composite Risk Scoring with Intent Verification.}
The access governance model proposed here extends the ABAC framework with two additional
components: a composite risk scoring function that evaluates access requests across multiple attribute
dimensions simultaneously, and an intent verification gate that activates when the composite score
exceeds a defined threshold, routing the request to an intent alignment check rather than to binary
approval or human escalation.

The composite risk scoring function evaluates each access request across four candidate attribute
dimensions that are proposed here as a structured starting point for empirical investigation rather
than as a validated specification. The first dimension is \textbf{data sensitivity}, classifying the data assets
the requested action would touch on a defined sensitivity scale from public through internal,
confidential, restricted, and critical, reflecting that an agent accessing canteen menu data presents
categorically different risk than one accessing engineering blueprints, proprietary model weights, or
personally identifiable information, even when both actions are performed under the same
credential. The second dimension is \textbf{action irreversibility}, distinguishing between read
operations-which do not modify data but whose disclosure effects may themselves be irreversible if
sensitive information reaches an unauthorized party-and write, delete, send, publish, or execute
operations, which may be irreversible in consequence and whose governance review cost is therefore
justified at lower probability thresholds. The third dimension is \textbf{contextual deviation}, measuring
the distance between the current request and the agent's documented behavioral baseline across
parameters including time, recipient, data volume, tool combination, and access pattern, drawing on
the behavioral monitoring architecture specified in MIGT Domain~III. The fourth dimension is
\textbf{task alignment}, measuring whether the requested tool-data combination is consistent with the task
specification under which the agent was deployed-the most direct operationalization of the
practitioner question identified in Section~1: are these tool and data combinations within the scope
of the given task?

Each dimension is scored on a normalized scale and combined into a composite risk score using a
weighted function whose weights are configurable by the organization's risk governance policy. The
weighting reflects organizational risk appetite: an organization handling classified government data
may weight data sensitivity heavily relative to contextual deviation, while a financial services
organization may weight action irreversibility most heavily given the regulatory consequences of
irreversible transactions executed without human review. The specific dimensions, scoring weights,
and threshold values proposed here are provisional and subject to empirical revision; their purpose
at this stage is to make the structure of the problem precise rather than to specify a production-ready
implementation.

\textbf{The Intent Verification Gate.}
A high composite risk score does not, in this model, automatically trigger human escalation. Human
escalation is expensive, introduces latency that may break agentic task completion, and is subject to
the cognitive overload risk identified in the OWASP Agentic Security Initiative's threat taxonomy
[OWASP, 2025], where adversaries deliberately generate high volumes of escalation requests to
exhaust human review capacity. The proposed model interposes an intent verification gate between
the composite risk score and the escalation decision.

The intent verification gate operates as a semantic alignment check: it evaluates whether the
requested tool-data combination is consistent with the natural language task specification under
which the agent was authorized, using a lightweight inference model operating on the task
description, the agent's current execution context, and the specific access request to produce an
alignment score. A critical architectural requirement is that this inference model must be isolated
from the agent's own reasoning context and executed outside the agent's trust boundary: because the
threat model includes prompt injection and goal manipulation, an alignment checker that draws on
the same context as the agent being evaluated is potentially subject to the same class of attacks it is
designed to detect. The integrity of the intent verification gate therefore depends on architectural
separation between the agent's reasoning layer and the verification layer, a requirement that
connects this proposal to the attestation architecture described in Section~6.3.1 and that represents an
additional open engineering problem the proposal inherits. A high composite risk score paired with a
high alignment score indicates a legitimate but sensitive access pattern that may warrant enhanced
logging and post-hoc review but not necessarily pre-execution escalation. A high composite risk
score paired with a low alignment score indicates a potential misalignment between execution and
authorization-the condition most consistent with prompt injection, goal manipulation, or
unauthorized task expansion-and warrants either automatic denial or human escalation depending
on the severity combination. This distinction operationalizes a governance principle that current
frameworks do not enforce: that the question is not only whether an action is within the agent's
technical permission envelope, but whether it is within the intent boundary of the task that justified
granting that envelope.

This approach is supported by recent work on intent-bound delegation in agentic systems, which
documents that binding actions to authenticated intent is the most promising mechanism for
preventing injected instructions from propagating across tools and services while preserving agent
functionality in legitimate task execution \cite{castro2026survey}. The Databricks AI Security
Framework v3.0 characterizes the access governance problem as a three-legged attack precondition:
privileged access combined with untrusted input processing and exfiltration capability. The intent
verification gate directly targets the second leg, distinguishing trusted from untrusted input by
evaluating alignment with authorized task intent rather than relying solely on input content filtering
\cite{databricks2026dasf}.

\textbf{Relationship to Existing MIGT Controls.}
The attribute-composite risk scoring model is not a replacement for the policy decision point
architecture specified in MIGT Domain~III but an extension of it: the composite risk score and
intent alignment score function as additional inputs to the policy decision point, enriching the
contextual rule evaluation with dimensions that current PDP implementations do not address. The
behavioral baseline established in Domain~III provides the contextual deviation dimension of the
composite score. The task specification recorded in the AI Identity Registry established in
Domain~I provides the intent boundary against which alignment is evaluated. The audit architecture
established in Domain~IV records composite risk scores and intent alignment decisions alongside
action logs, providing the forensic record necessary to reconstruct why a specific access decision
was made and whether the intent verification gate operated correctly. The model is therefore
architecturally dependent on the foundational governance capabilities of the preceding MIGT
domains and is not implementable in their absence.

\textbf{Implementation Maturity and Research Status.}
The attribute-composite risk scoring model is proposed here as a governance design concept
grounded in existing technical capabilities rather than a fully validated implementation. The
component technologies-ABAC policy engines, behavioral baseline monitoring, and LLM-based
semantic alignment checking-are each available in current enterprise security tooling. Their
integration into a composite risk scoring function with an intent verification gate represents a
research and development agenda rather than a deployable product as of March~2026. The
practitioner literature confirms that intent-bound access governance is recognized as an open
problem in agentic AI security \cite{castro2026survey, owasp2025agentic}, and the model proposed here is
offered as a contribution to that research agenda. Empirical validation through controlled
deployment studies-of the kind that Section~8.4 identifies as a priority future research direction for
the MIGT implementation roadmap more broadly-is required before the model's specific parameter
choices, scoring weights, and threshold values can be specified with confidence.

\subsection{MIGT Domain IV: Accountability Attribution and Audit Architecture}

\textbf{Purpose:} Ensure every AI agent action is attributable to a specific, cryptographically verified
identity, that complete, tamper-evident audit records exist for every significant AI action, and that
accountability for AI-caused harm can be clearly assigned within the governance framework.
\textbf{Gap addressed:} the Audit Trail Insufficiency, Diffuse Accountability for AI Harm, and Human
Oversight Nominalism risks in AIRT Domain~V. Required capabilities include:

\begin{enumerate}
    \item \textbf{Unique Agent Identity as Accountability Anchor}, where every AI agent is assigned a
    unique, persistent identifier that is cryptographically linked to the agent's model version and
    deployment configuration, recorded in the AI Identity Registry, appears in every audit log
    entry, and is maintained across the agent's lifecycle \cite{ref38};

    \item \textbf{Tamper-Evident Action Audit Trail}, where every significant AI agent action is logged with
    context-including identity, policy, data classification, and session scope-in a format that
    enables forensic reconstruction. Every credential request, approval, and usage event must
    generate detailed logs feeding into observability platforms that enable anomaly detection,
    compliance reporting, and forensic investigations \cite{ref71};

    \item \textbf{Designated Human Accountability Owner}, where every AI identity has a designated human
    owner bearing genuine accountability, addressing the Human Oversight Nominalism risk in
    substance, not merely form \cite{ref38}\cite{ref39}; and

    \item \textbf{Incident Response and Accountability Attribution Protocol} providing a defined process for
    identifying the specific agent involved, reconstructing the action sequence, determining
    contributing governance failures, and assigning accountability to the appropriate human owner
    and organizational level.
\end{enumerate}

\noindent\textbf{Regulatory alignment:} EU AI Act Articles~12 and~17 (record-keeping and quality management
systems), China CSL risk monitoring and safety assessments \cite{ref54}, NIST AI RMF GOVERN and
MANAGE functions, SOX AI transparency requirements.

\subsection{MIGT Domain V: Supply Chain and Model Provenance Governance}

\textbf{Purpose:} Ensure AI models and components deployed in enterprise environments have verifiable
provenance-including documented training data, model lineage, component dependencies, and
modification history-and that supply chain governance controls reduce the risk of malicious model
injection, training data poisoning, and third-party component compromise. \textbf{Gap addressed:} AIRT
Domain~VI risks. The ACM survey \cite{ref34} identifies training data poisoning as a mechanism for
embedding harmful behaviors that standard evaluation does not detect, making supply chain
governance a prerequisite for meaningful AI identity governance. Required capabilities include:

\begin{enumerate}
    \item \textbf{AI Bill of Materials (AIBOM)}, where every AI model deployed in enterprise environments
    has a documented AIBOM specifying model architecture, training data sources and governance
    status, fine-tuning history, dependency components and versions, and modification records
    \cite{ref63};

    \item \textbf{Model Integrity Verification} using cryptographic hash verification of model parameters
    with FIPS-approved SHA-3 family hash functions \cite{ref63};

    \item \textbf{Third-Party Component Governance} applying vendor risk management to AI agent
    frameworks, tool integrations, plugin ecosystems, and API dependencies, with specific
    attention to AI-specific vulnerabilities identified in the ACM survey \cite{ref34}; and

    \item \textbf{GPAI Model Compliance Documentation} for AI agents built on GPAI models subject to
    EU AI Act Article~53 obligations \cite{ref46}.
\end{enumerate}

\noindent\textbf{Regulatory alignment:} EU AI Act Articles~53 and~55 (GPAI model obligations), U.S.\ AI Action
Plan supply chain security requirements \cite{ref51}, China CSL AI safety assessments \cite{ref54}, NIST AI
RMF MAP function.

\subsection{MIGT Domain VI: Regulatory Alignment and Cross-Jurisdictional Coordination}

\textbf{Purpose:} Provide a structured framework for mapping enterprise AI identity governance obligations
across multiple regulatory jurisdictions simultaneously, identifying cross-jurisdictional conflicts,
and providing governance mechanisms for managing those conflicts at the enterprise program level.
\textbf{Gap addressed:} the Regulatory Fragmentation Problem (Section~1.2, Gap~5), the three fault lines
documented in Section~4.6, and the Cross-Jurisdictional Regulatory Weaponization risk in AIRT
Domain~VII. Required capabilities include:

\begin{enumerate}
    \item A \textbf{Jurisdiction Mapping Matrix}-a structured, maintained mapping of AI identity governance
    obligations applicable to the organization's operational jurisdictions against the governance
    capabilities in MIGT Domains~I through~V (see Table~\ref{tab:regulatory} below);

    \item A \textbf{Cross-Jurisdictional Conflict Registry}-a maintained registry of identified conflicts
    documenting the specific conflicting requirements, the jurisdictions involved, the nature of
    the conflict, the organization's current management approach, and the residual risk;

    \item \textbf{Jurisdictional Tiering of AI Deployments}, classifying AI deployments by jurisdictional
    exposure with governance controls applied proportional to each deployment's exposure; and

    \item \textbf{Regulatory Velocity Monitoring}-a structured process for monitoring regulatory
    developments and updating the Jurisdiction Mapping Matrix and Conflict Registry as the
    landscape evolves, given the rapid and asynchronous pace of regulatory change documented in
    Section~4.6.
\end{enumerate}


\definecolor{highconflictred}{RGB}{200, 40, 40}
\definecolor{moderateorange}{RGB}{190, 100, 20}
\definecolor{lowgreen}{RGB}{40, 140, 60}

\newcommand{\conflictHigh}[1]{%
  \textcolor{highconflictred}{\textbf{HIGH CONFLICT:} #1}}
\newcommand{\conflictModerate}[1]{%
  \textcolor{moderateorange}{\textbf{Moderate conflict:} #1}}
\newcommand{\conflictLow}[1]{%
  \textcolor{lowgreen}{\textbf{Low conflict:} #1}}

\newcolumntype{Y}{>{\RaggedRight\arraybackslash}X}



\setlength{\extrarowheight}{2pt}

\begin{landscape}
\begin{center}
\begin{longtable}{%
  >{\RaggedRight\arraybackslash}p{0.18\linewidth}
  >{\RaggedRight\arraybackslash}p{0.16\linewidth}
  >{\RaggedRight\arraybackslash}p{0.16\linewidth}
  >{\RaggedRight\arraybackslash}p{0.16\linewidth}
  >{\RaggedRight\arraybackslash}p{0.22\linewidth}}

\caption{MIGT Regulatory Alignment Matrix. HIGH CONFLICT entries
indicate irreconcilable cross-jurisdictional obligations requiring
active management through the MIGT Domain~VI Conflict Registry.}
\label{tab:regulatory} \\

\toprule
\textbf{MIGT Domain}
  & \textbf{EU AI Act Obligation}
  & \textbf{U.S.\ (NIST AI RMF + State)}
  & \textbf{China (CSL/DSL/GenAI)}
  & \textbf{Conflict Assessment} \\
\midrule
\endfirsthead

\toprule
\textbf{MIGT Domain}
  & \textbf{EU AI Act Obligation}
  & \textbf{U.S.\ (NIST AI RMF + State)}
  & \textbf{China (CSL/DSL/GenAI)}
  & \textbf{Conflict Assessment} \\
\midrule
\endhead

\midrule
\multicolumn{5}{r}{\emph{Continued on the next page}} \\
\endfoot

\bottomrule
\multicolumn{5}{l}{%
  \footnotesize\textit{Sources:} EU AI Act
  \cite{ref46}\cite{ref47}\cite{ref48};
  NIST AI RMF \cite{ref15}\cite{ref26}\cite{ref44};
  China CSL/GenAI \cite{ref54}\cite{ref55}\cite{ref56};
  Cross-jurisdictional analysis
  \cite{ref30}\cite{ref31}\cite{ref95}\cite{ref96}.} \\
\endlastfoot

I.\ Identity Lifecycle
  & Art.~9 risk mgmt; Art.~17 QMS documentation
  & NIST AI RMF GOVERN function; state transparency laws
  & CSL (Jan.\ 2026) AI compliance; risk monitoring requirements
  & \conflictLow{all three frameworks require lifecycle documentation
    and ownership assignment} \\[3pt]

II.\ Cryptographic Identity
  & Art.~15 cybersecurity by design; Art.~9 risk management
  & NIST SP~800-207 zero trust architecture; NCCoE AI agent
    identity initiative
  & CSL network security requirements; algorithm security standards
  & \conflictLow{technical standard convergence toward cryptographic
    workload identity} \\[3pt]

III.\ Dynamic Access
  & Art.~14 human oversight requirements; Art.~9 risk management
  & NIST AI RMF MEASURE function; CISA zero trust maturity model
  & Algorithm Recommendation Measures behavioral monitoring
    requirements
  & \conflictModerate{EU requires human override capability; China
    requires state content alignment in behavioral monitoring} \\[3pt]

IV.\ Audit \& Accountability
  & Art.~12 record-keeping; Art.~17 quality management systems
  & NIST AI RMF MANAGE function; SOX AI transparency requirements
  & CSL AI risk monitoring; safety assessment documentation
  & \conflictModerate{log retention periods and audit access rights
    may conflict with GDPR data minimization obligations} \\[3pt]

V.\ Supply Chain
  & Arts.~53--55 GPAI model obligations; Art.~9 risk management
  & NIST AI RMF MAP function; SBOM requirements for federal
    procurement
  & CSL AI safety assessments; CAC algorithm filing and transparency
    requirements
  & \conflictHigh{EU and U.S.\ trade secret protection obligations
    directly incompatible with Chinese algorithm transparency filing
    requirements for AI systems} \\[3pt]

VI.\ Cross-Jurisdictional Coordination
  & Art.~4 extraterritorial application; Digital Omnibus obligations
  & EO~14365 federal preemption signals; state law uncertainty
    pending court resolution
  & CSL extraterritorial reach effective Jan.\ 1, 2026; national
    security disclosure authority
  & \conflictHigh{incompatible extraterritorial claims from three
    jurisdictions; no international resolution mechanism exists above
    enterprise program level} \\

\end{longtable}
\end{center}
\end{landscape}

\setlength{\extrarowheight}{0pt}

\subsection{MIGT and the State Actor Threat Model}

The six governance domains above address the framework gap and the regulatory gap. The MIGT
additionally incorporates a foreign state actor threat model as an overlay across all six domains,
adding specific governance requirements triggered when threat intelligence indicates state actor
targeting. These requirements include: enhanced credential security for PAM-adjacent identities
given Silk Typhoon's documented targeting of PAM provider credentials \cite{ref58}\cite{ref59}; API key
inventory and foreign exposure assessment for every machine credential; supply chain foreign
provenance review specifically assessing state actor risk in AI model and component procurement
in addition to standard vendor risk management; and cross-jurisdictional data access governance for
organizations operating under Chinese CSL jurisdiction, with specific conflict registry entries for
the CSL's extraterritorial access provisions effective January~1, 2026 \cite{ref54}.

\subsection{The MIGT as an Integrated Response to Five Structural Gaps}

The six domains of the MIGT map directly onto the five structural gaps identified in Section~1.2.
Domain~I's AI Identity Registry and lifecycle governance processes address Gap~1 (Identity
Continuum) by applying governance discipline regardless of where on the human-machine spectrum
an AI identity falls. Domain~III's JIT provisioning and policy decision architecture address

\section{Implementation Guidance: Translating the MIGT into Enterprise Programs}

The MIGT presented in Section~6 defines what governance capabilities enterprise organizations
need. This section addresses how to build them, providing a structured, phased implementation
roadmap that translates the MIGT's six domains into actionable program steps, sequenced by
dependency and risk priority. The roadmap is grounded in the Identity Working Group's
practitioner-validated maturity model \cite{ref73}, the Permiso NHI maturity progression \cite{ref74},
the KuppingerCole IAM governance roadmap methodology \cite{ref75}, the OASIS NHI management
guide \cite{ref76}, and the enterprise AI security roadmap validated by CISO practitioners \cite{ref77}. The
roadmap has not been tested through prospective longitudinal implementation studies; that
limitation, and the research agenda it generates, is addressed in Section~8.3. It is offered as a
practitioner-grounded starting framework that future implementation studies should validate,
refine, and extend.

\subsection{Pre-Implementation: The Current State Assessment}

Before beginning MIGT implementation, organizations must establish an honest baseline of their
current AI identity governance posture. The current state assessment answers five questions:
\begin{enumerate}
  \item \textbf{What AI identities exist?} Conduct a discovery sweep across all environments,
    targeting source code and CI/CD workflows first, where 57\% and 26\% of exposed secrets
    respectively originate \cite{ref45};
  \item \textbf{What do those identities have access to?} Audit effective permissions, not just
    formally assigned roles, with specific attention to Super NHIs, given that 1 in 20 AWS
    machine identities carries full administrator privileges \cite{ref80};
  \item \textbf{Who owns them?} Orphaned identity prevalence is the primary baseline metric for
    remediation tracking;
  \item \textbf{What is the current credential hygiene posture?} Document the prevalence of static
    API keys, long-lived service account passwords, hardcoded secrets, and shared credentials;
    and
  \item \textbf{What regulatory frameworks apply?} Use the MIGT Domain~VI Jurisdiction Mapping
    Matrix to determine applicable frameworks and begin jurisdictional exposure mapping.
\end{enumerate}

The output is a Current State Report providing the before-measurement for demonstrating risk
reduction as the program progresses.

\subsection{Phase 1: Foundation (Months 1--6)}

\textbf{Objective:} Establish the foundational governance infrastructure that all subsequent phases
depend on, and deliver measurable reduction in the highest-severity risks identified in the current
state assessment. The Identity Working Group \cite{ref73} identifies unified architecture and visibility
as the essential first step; Permiso's maturity model \cite{ref74} confirms that NHI visibility grows
from less than 10\% to over 99\% through foundation-level implementation, representing the largest
single risk reduction step in the entire maturity progression.

Phase 1 deliverables:
\begin{enumerate}
  \item \textbf{AI Identity Registry,} initially populated with unique identifier, identity type,
    business purpose, provisioning date, access grants, associated systems, and designated human
    owner, with a specific shadow AI category for identities discovered outside formal provisioning
    processes (repositories with AI assistance enabled show a 40\% higher incidence of secret
    leaks \cite{ref73});
  \item \textbf{Ownership Assignment Campaign,} designating a human accountability owner for
    every AI identity through engagement with business unit leaders and application owners, with
    a triage process for orphaned identities;
  \item \textbf{Critical Credential Remediation,} prioritized as Tier~1 (immediate) for hardcoded
    secrets in source code and CI/CD pipelines, PAM-adjacent credentials identified in Silk
    Typhoon threat intelligence \cite{ref58}, credentials with administrative scope, and credentials
    with no rotation history exceeding 12 months; Tier~2 (30--60 days) for API keys with broad
    scope and no usage monitoring; and Tier~3 (60--90 days) for remaining static long-lived
    credentials. The OASIS NHI management guide \cite{ref76} provides the most operationally
    detailed guidance for credential rotation in high-debt environments without causing service
    disruption; and
  \item \textbf{Authorized Provisioning Process,} establishing a lightweight initial version creating
    a gate that prevents future shadow AI identity accumulation.
\end{enumerate}

Phase 1 success metrics: AI identity registry populated with greater than 80\% completeness
confidence; designated owner assigned to greater than 90\% of active AI identities; Tier~1
credential remediation 100\% complete; authorized provisioning process operational for all new AI
identity requests.

\subsection{Phase 2: Hardening (Months 6--12)}

\textbf{Objective:} Replace static credential architecture with cryptographic identity and ephemeral
credentials, implement JIT access provisioning for high-risk AI identities, establish behavioral
baselines enabling anomaly detection, and begin EU AI Act and China CSL compliance
documentation. The EU AI Act August 2026 enforcement date creates a hard external deadline for
EU-exposed organizations.

Phase 2 deliverables:
\begin{enumerate}
  \item \textbf{SPIFFE/SPIRE Workload Identity Deployment} \cite{ref68} for the highest-risk AI
    identity population, beginning with PAM-adjacent identities, followed by identities with
    access to sensitive data systems, followed by remaining high-privilege identities. SPIFFE
    provides cryptographically verifiable IDs tied to workloads, not people, making them ideal
    for AI agents and other non-human entities \cite{ref69};
  \item \textbf{Ephemeral Credential Architecture} for Tier~1 high-risk identities, with Zero
    Standing Privilege as the target architecture where agents never hold a credential at rest
    \cite{ref70};
  \item \textbf{JIT Access Provisioning} pilot for a selected population of AI agents with currently
    broad standing access to sensitive systems, demonstrating that access can be granted at
    runtime, scoped to specific tasks, and automatically revoked upon completion \cite{ref71};
  \item \textbf{Behavioral Baseline Establishment} for each AI identity derived from historical log
    analysis; and
  \item \textbf{Compliance Documentation,} initial version mapping implemented controls to EU AI
    Act and China CSL obligations using the MIGT Domain~VI Jurisdiction Mapping Matrix.
\end{enumerate}

Phase 2 success metrics: SPIFFE/SPIRE deployed for greater than 50\% of Tier~1 AI identities;
ephemeral credentials operational for all PAM-adjacent AI identities; JIT provisioning pilot
operational with documented risk reduction measurement; behavioral baselines documented for
greater than 80\% of registered AI identities; EU AI Act compliance gap assessment completed with
remediation roadmap.

\subsection{Phase 3: Integration (Months 12--24)}

\textbf{Objective:} Extend cryptographic identity and ephemeral credentials to the full AI identity
population, deploy tamper-evident audit architecture, complete supply chain governance for
high-risk AI models, and achieve full cross-jurisdictional regulatory alignment documentation.

Phase 3 deliverables:
\begin{enumerate}
  \item \textbf{Full SPIFFE/SPIRE rollout} across the complete AI identity population, retiring all
    remaining static long-lived credentials, with ARIA model implementation \cite{ref66} for
    multi-agent architectures enabling fine-grained accountability across complex multi-agent
    workflows;
  \item \textbf{Tamper-Evident Audit Architecture} deploying complete logging of all significant AI
    agent actions enriched with context and integrated with SIEM and SOAR platforms \cite{ref71};
  \item \textbf{AIBOM Implementation} for all production AI models with model integrity
    verification using cryptographic hash verification of model parameters \cite{ref63};
  \item \textbf{Cross-Jurisdictional Conflict Registry,} fully populated with documented management
    approaches for identified irreconcilable conflicts, validated by legal counsel, particularly for
    organizations operating across U.S.-China boundaries where the algorithm transparency
    conflict identified in the regulatory matrix has no regulatory resolution; and
  \item \textbf{AI-Specific Access Certification} using automated behavioral analysis combined with
    designated owner attestation.
\end{enumerate}

Phase 3 success metrics: cryptographic identity deployed across greater than 95\% of AI identity
population; zero active static long-lived credentials for Tier~1 identities; tamper-evident audit
architecture operational with SIEM integration; AIBOM complete for all production AI models; EU
AI Act compliance documentation complete and legally reviewed; Cross-Jurisdictional Conflict
Registry fully populated.

\subsection{Phase 4: Optimization (Month 24+)}

\textbf{Objective:} Advance the AI identity governance program from human-speed reactive
management to machine-speed proactive governance.

Phase 4 deliverables:
\begin{enumerate}
  \item \textbf{Policy Decision Point Automation,} implementing automated contextual rule
    evaluation without requiring human review for routine decisions, with human review triggered
    automatically for high-risk deviations and state-actor threat indicators \cite{ref64};
  \item \textbf{Threat Intelligence Integration,} incorporating CISA advisories, Microsoft Threat
    Intelligence bulletins \cite{ref58}, and FBI Cyber Division notifications into anomaly detection
    and policy decision architecture, with Silk Typhoon-specific behavioral detection rules for
    anomalous access to PAM-adjacent identities;
  \item \textbf{Maturity-Based Autonomy Progression Management} \cite{ref64}, continuously
    assessing each AI agent's compliance and updating autonomy tiers accordingly; and
  \item \textbf{Regulatory Velocity Monitoring,} establishing a structured process for monitoring EU
    AI Act implementation guidance, NIST NCCoE AI agent identity standards \cite{ref19}\cite{ref44},
    Chinese CSL enforcement guidance, and U.S.\ state AI law developments.
\end{enumerate}

\subsection{Implementation Guidance for Specific Organizational Contexts}

Organizations with significant identity debt should execute a pre-Phase 1 triage step using threat
intelligence to prioritize the highest-risk identity subset for immediate action before full inventory
is complete: PAM-adjacent credentials, identities with administrative scope, and credentials in
externally accessible systems. The OASIS NHI management guide \cite{ref76} provides the most
detailed practitioner guidance for credential rotation in high-debt environments without causing
service disruption. Organizations subject to all three primary regulatory frameworks-the EU AI
Act, NIST AI RMF, and Chinese CSL-should initiate the Cross-Jurisdictional Conflict Registry
during the current state assessment phase rather than Phase~3, and engage legal counsel with
specific expertise in all three frameworks before Phase~2 compliance documentation begins. The
August 2026 EU AI Act enforcement date creates a hard external deadline that must drive Phase~2
and Phase~3 scheduling for EU-exposed organizations regardless of other program priorities.

\subsection{Measuring Program Effectiveness}

MIGT implementation programs should track five categories of metrics:
\begin{itemize}
  \item \textbf{Coverage Metrics:} percentage of AI identities in registry, with designated owners,
    with cryptographic identity, and with behavioral baselines, measuring the breadth of
    governance program reach.
  \item \textbf{Hygiene Metrics:} percentage of credentials rotated within policy period, percentage
    JIT-provisioned, and number of active static long-lived credentials (target zero for Tier~1),
    measuring credential governance quality.
  \item \textbf{Behavioral Metrics:} number of behavioral baseline deviations detected, mean time
    to investigate anomalies, and percentage attributed to known threat actor patterns, measuring
    the effectiveness of behavioral monitoring.
  \item \textbf{Accountability Metrics:} percentage of AI incidents with complete audit trail
    reconstruction, mean time to accountability attribution, and number of AI identities with no
    designated owner (target zero), measuring the strength of accountability architecture.
  \item \textbf{Compliance Metrics:} percentage of applicable regulatory obligations mapped to
    governance controls, number of open cross-jurisdictional conflicts without documented
    management approaches, and status of EU AI Act compliance documentation against the
    August 2026 enforcement date.
\end{itemize}

\section{Conclusion: Governing the Machine - Four Claims, an Inflection Point, and a Research Agenda}

The three questions that most organizations cannot answer about their AI systems in 2026-What
can they access? Who is responsible when they cause harm? And does governing them in one
country create liability in another?-eight sections later, the empirical record, the risk taxonomy,
the governance framework, and the regulatory analysis collectively demonstrate that these questions
are not merely unanswered. They are, in most enterprise environments, not yet being seriously
asked. Here, the concluding section restates four original claims against the evidence developed
across the preceding sections, acknowledges limitations with the intellectual honesty that peer
review requires, and outlines the future research agenda that the MIGT and AIRT open.

\subsection{Restating the Four Claims}

\textbf{Claim 1:} AI identity risk is a distinct risk category requiring its own taxonomy. Section~2
documents machine identities outnumbering humans by 144 to 1 \cite{ref45}, governance capabilities
covering fewer than half the total machine identity population \cite{ref3}, and ungoverned permissions
growing from 5\% to 28\% of enterprise total in a single year \cite{ref80}. The CrowdStrike outage
(\$5.4 to \$10 billion in losses \cite{ref84}\cite{ref85}), the tj-actions supply chain compromise (23,000+
repositories \cite{ref45}), and the ongoing Typhoon campaigns collectively demonstrate that machine
identity governance failure is producing harm at national scale. The State of AI Agent Security 2026
report confirms that 88\% of organizations experienced AI agent security incidents yet only 22\%
treat AI agents as independent, identity-bearing entities \cite{ref89}. The AIRT's 37 sub-categories
across eight domains constitute the first comprehensive enumeration of the risk category these
numbers describe.

\textbf{Claim 2:} The governance framework gap, the regulatory gap, and the coordination gap must
be addressed simultaneously. The agentic AI governance frameworks of Pandey \cite{ref28}, Joshi
\cite{ref29}, and Shavit et al.\ \cite{ref38} address governance broadly without identity-specific grounding.
The NIST AI RMF \cite{ref15} provides governance function structure without identity-specific risk
enumeration. Cross-jurisdictional analyses \cite{ref30}\cite{ref31} document regulatory divergence without
enterprise-level resolution mechanisms. The Cloud Security Alliance survey of 285 IT and security
professionals confirms that less than half feel confident they could pass a compliance review focused
on agent behavior \cite{ref90}-the operational consequence of addressing these gaps in isolation. The
MIGT addresses all three gaps simultaneously.

\textbf{Claim 3:} Foreign state actors have already operationalized AI identity as an attack and
influence vector, and enterprise IAM programs have no adequate threat model for this. Silk
Typhoon's targeting of PAM provider credentials \cite{ref58}\cite{ref59}, Salt Typhoon's compromise of nine
major U.S.\ telecommunications providers still active as of February 2026 \cite{ref60}\cite{ref62}, Volt
Typhoon's pre-positioning within critical infrastructure \cite{ref61}, and North Korean AI-enhanced
identity fraud \cite{ref10} collectively establish that foreign state exploitation of AI identity governance
gaps is a present operational reality. The Lumos Identity at a Breaking Point report confirms that
96\% of organizations experienced identity-related security incidents \cite{ref92}. No existing IAM
governance framework reviewed in Section~3 incorporates a foreign state actor threat model.

\textbf{Claim 4:} The absence of inter-jurisdictional coordination in AI governance is itself a
governance risk. Section~4 documents five frameworks and three structural fault lines producing
irreconcilable cross-jurisdictional conflicts. The January 22, 2026 publication of Singapore's Model
AI Governance Framework for Agentic AI \cite{ref91} confirms the directional trend: major jurisdictions
continue to develop independent AI governance frameworks that enterprise organizations must
simultaneously satisfy, without any international coordination mechanism to manage the resulting
conflicts. Schmitt \cite{ref96} documented the fragmentation of this governance regime as early as 2021,
noting the risk of geopolitical encroachment into technical governance domains. Perboli et al.\
\cite{ref95} confirm that by 2025 regulatory arbitrage and normative fragmentation have materialized as
predicted. The MIGT's Regulatory Alignment Domain and the Regulatory Matrix constitute the first
structured enterprise-level response to this coordination failure in the published literature.

\subsection{The Inflection Point}

The four claims describe not merely a governance gap but a governance inflection point. The
governance deficit-whereby organizations struggle to govern machine-speed identities using
human-speed manual processes-combined with agentic risk, where AI agents now act with
administrative privileges that often exceed those of their human creators, defines the current
moment. IANS Research characterizes 2026 as the year AI agents are creating an identity security
crisis, documenting that the window to get NHI governance right is narrowing as attackers refine
tactics faster than governance programs can adapt \cite{ref94}. At the same time, the achievability
dimension is established by converging standards, tooling, and regulatory pressure: SPIFFE/SPIRE
has graduated from the CNCF with production-ready cryptographic workload identity capability
\cite{ref68}, the NCCoE has initiated the AI agent identity standardization project \cite{ref44}\cite{ref19}, and
the EU AI Act's August 2026 enforcement date is driving governance investment across global
organizations. Keyfactor research finds that 85\% of cybersecurity professionals expect digital
identities for agents will be as common as human and machine identities within five years, while
86\% say that without unique and dynamic digital identities, AI agents cannot be fully trusted. The
MIGT is proposed as a governance framework for this inflection point.

\subsection{Limitations}

Four limitations are material and should be acknowledged explicitly. First, AIRT severity ratings
have not been empirically validated through large-scale enterprise security incident data. Future
research should test AIRT classifications against incident databases, including the VERIS
Community Database, the Privacy Rights Clearinghouse breach database, and AI-specific incident
repositories that the EU AI Act and China CSL reporting requirements will, over time, generate.
Second, the MIGT implementation roadmap is grounded in practitioner-validated sources but has
not been tested through prospective longitudinal implementation studies. Enterprise implementation
complexity varies significantly with organizational size, technical debt, regulatory exposure, and
governance maturity. Third, the cross-jurisdictional analysis in Sections 4 and 6 addresses three
primary regulatory jurisdictions. The January 2026 Singapore MGF \cite{ref91} demonstrates that the
challenge is expanding to additional jurisdictions not fully addressed in this paper, and the UK AI
Safety Institute framework represents a fifth major jurisdiction with independent AI governance
obligations following Brexit. The MIGT's Domain~VI framework is extensible to additional
jurisdictions through the Jurisdiction Mapping Matrix mechanism, and the Wang et al.\ \cite{ref99}
quantitative analysis of China, EU, and U.S.\ policy frameworks provides a methodological
foundation for extending that analysis to additional jurisdictions in future work. The specific
conflict analysis for Singapore, the UK, and other emerging national frameworks is beyond this
paper's scope. Fourth, the taxonomy is point-in-time as of March 2026 and will require revision as
new attack techniques are documented, particularly in the prompt injection and state actor domains.

\subsection{Future Research Directions}

Six research directions are identified as particularly productive. First, empirical validation and
calibration of the AIRT through enterprise security incident data, requiring research partnerships
with enterprise security operations centers, insurance carriers with AI-specific policy portfolios,
and regulatory bodies with incident reporting access. Second, MIGT implementation studies and
maturity model development through longitudinal case studies across organizations of different
sizes, regulatory exposures, and technical debt levels, with the Omada State of Identity Governance
Report 2026 \cite{ref21} providing a useful baseline measurement against which future implementation
studies should reference. Third, the Model Context Protocol (MCP) as an emerging identity
governance frontier: MCP tokens, credentials, and access rules are becoming primary targets
because they enable agents to operate within critical systems at machine speed, and most
organizations lack visibility into this layer. Future work should extend the AIRT to include
MCP-specific risk sub-categories and the MIGT to include MCP-specific governance controls.
Fourth, cross-jurisdictional conflict resolution mechanisms, investigating whether international
instruments-including bilateral regulatory equivalence agreements, mutual recognition frameworks,
or sector-specific international governance bodies-could provide resolution mechanisms for the
specific irreconcilable conflicts identified in Section~4.6. The Carnegie Endowment's analysis of
international AI governance regime complexity \cite{ref43} and Schmitt's foundational regime mapping
\cite{ref96} both identify this as a productive area for international relations and governance
scholarship. Fifth, the OT and consumer AI extension of the AIRT and MIGT to operational
technology environments, where the consequences of governance failure may be physical as well as
informational, and to consumer-facing AI deployments where the governance requirements differ
meaningfully from the enterprise context analyzed. Sixth, the non-human identity accountability
attribution problem in international law: when a foreign state actor exploits an ungoverned machine
credential to cause harm across international borders, the Tallinn Manual tradition \cite{ref93} has not
yet developed adequate doctrine for the AI identity context. This intersection of international
humanitarian law, state responsibility doctrine, and AI identity governance is an underexplored area
the foreign state actor threat model opens for future research, and one with direct policy implications
for every government currently developing AI governance frameworks.

\subsection{A Closing Observation}

The governance vacuum at the intersection of AI and identity is not a future problem. It is a present
crisis, documented in the preceding eight sections with data, incidents, regulatory analysis, and
threat intelligence that collectively leave no room for the comfortable assumption that existing
frameworks are adequate or that incremental improvement will close the gap in time.

AI agents are no longer experimental. They are production infrastructure. 81\% of technical teams
have moved past the planning phase into active testing or production deployment \cite{ref89}. Yet only
14.4\% report that AI agents go live with full security and IT approval \cite{ref89}, meaning that the
overwhelming majority of AI agents entering enterprise production environments today are doing so
without adequate governance review, without cryptographic identity, without behavioral monitoring,
and without designated human accountability owners.

The MIGT does not solve this problem. No taxonomy or framework solves an operational
governance problem. That requires investment, organizational commitment, and sustained
execution. What the MIGT provides is the structured intellectual foundation that makes solving the
problem possible: a comprehensive risk classification, a governance framework designed for the
specific characteristics of AI identity risk, an implementation roadmap grounded in practitioner
experience, and a cross-jurisdictional regulatory alignment structure that gives global organizations
the tools to navigate an increasingly fragmented governance landscape.

The question posed-who governs the machine-does not have a universal answer. It has an
organizational answer, a regulatory answer, and a geopolitical answer, and they are not the same
answer in any two jurisdictions. But in every jurisdiction, in every enterprise, and under every
regulatory framework, the answer begins the same way: with an inventory, an owner, a credential
policy, and an audit trail. With knowing what you have, who is responsible for it, and what it is
doing.

That is where governance begins. The MIGT provides the framework for everything that follows.

\newpage
\bibliographystyle{unsrt}
\bibliography{references_raw}

@misc{ref1,
  author = {Peck, T.},
  title = {How {AI} is reshaping identity governance for {CISOs} and {CIOs}},
  howpublished = {CyberArk Blog},
  year = {2025},
  month = nov,
  note = {Available: \url{https://www.cyberark.com/resources/blog/how-ai-is-reshaping-identity-governance-for-cisos-and-cios}},
}

@misc{ref2,
  author = {Gyure, W. and Johnson, K.},
  title = {The Human-Machine Identity Blur: A Unified Framework for Cybersecurity Risk Management in 2025},
  howpublished = {arXiv:2503.18255 [preprint]},
  year = {2025},
  month = mar,
  note = {Available: \url{https://arxiv.org/abs/2503.18255}},
}

@techreport{ref3,
  author = {{IBM Security}},
  title = {{IBM} Security {X-Force} Threat Intelligence Index 2025},
  institution = {IBM},
  year = {2025},
  note = {Available: \url{https://www.ibm.com/reports/threat-intelligence}},
}

@misc{ref4,
  author = {{Delinea}},
  title = {Five identity-driven shifts reshaping enterprise security in 2026},
  howpublished = {Help Net Security},
  year = {2025},
  month = dec,
  note = {Available: \url{https://www.helpnetsecurity.com/2025/12/24/five-identity-driven-shifts-reshaping-enterprise-security-in-2026/}},
}

@misc{ref5,
  author = {Chin, S. and Lester, P.},
  title = {Protecting our edge: Trade secrets and the global {AI} arms race},
  howpublished = {CSIS},
  year = {2025},
  month = dec,
  note = {Available: \url{https://www.csis.org/analysis/protecting-our-edge-trade-secrets-and-global-ai-arms-race}},
}

@techreport{ref6,
  author = {{Fortune Business Insights}},
  title = {Identity and Access Management Market Size, Share, and Industry Analysis --- Forecast 2024--2032},
  institution = {Fortune Business Insights},
  year = {2024},
  month = sep,
  note = {Available: \url{https://www.fortunebusinessinsights.com/industry-reports/identity-and-access-management-market-100373}},
}

@misc{ref7,
  author = {{Identity Defined Security Alliance}},
  title = {Identity and Access Management in the {AI} Era: 2025 Guide},
  howpublished = {IDSA},
  year = {2025},
  note = {Available: \url{https://www.idsalliance.org/blog/identity-and-access-management-in-the-ai-era-2025-guide/}},
}

@inproceedings{ref8,
  author = {Chan, A. and Salganik, R. and Markelius, A. and others},
  title = {Harms from Increasingly Agentic Algorithmic Systems},
  booktitle = {Proc. 2023 ACM Conf. Fairness, Accountability, Transparency (FAccT 2023)},
  address = {Chicago, IL},
  year = {2023},
  month = jun,
  doi = {10.1145/3593013.3594033},
  note = {Available: \url{https://dl.acm.org/doi/10.1145/3593013.3594033}},
}

@misc{ref9,
  author = {{Lawfare Media}},
  title = {Algorithmic Foreign Influence: Rethinking Sovereignty in the Age of {AI}},
  howpublished = {Lawfare},
  year = {2025},
  month = aug,
  note = {Available: \url{https://www.lawfaremedia.org/article/algorithmic-foreign-influence--rethinking-sovereignty-in-the-age-of-ai}},
}

@misc{ref10,
  author = {{NYU Journal of Intellectual Property and Entertainment Law}},
  title = {Artificial Intelligence and State-Sponsored Cyber Espionage},
  howpublished = {JIPEL},
  year = {2025},
  note = {Available: \url{https://jipel.law.nyu.edu/artificial-intelligence-and-state-sponsored-cyber-espionage/}},
}

@misc{ref11,
  author = {{DFRLab}},
  title = {{FIMI} 101: Foreign Information Manipulation and Interference Targeting the 2024 {US} General Election},
  howpublished = {Atlantic Council},
  year = {2024},
  month = sep,
  note = {Available: \url{https://dfrlab.org/2024/09/26/fimi-101/}},
}

@misc{ref12,
  author = {{Cloud Security Alliance}},
  title = {{AI} and Privacy: Shifting from 2024 to 2025},
  howpublished = {CSA Blog},
  year = {2025},
  month = apr,
  note = {Available: \url{https://cloudsecurityalliance.org/blog/2025/04/22/ai-and-privacy-2024-to-2025-embracing-the-future-of-global-legal-developments}},
}

@misc{ref13,
  author = {{MIT AI Risk Repository}},
  title = {{AI} Risk Repository (v4)},
  howpublished = {Massachusetts Institute of Technology},
  year = {2025},
  month = dec,
  note = {Available: \url{https://airisk.mit.edu}},
}

@misc{ref14,
  author = {{OWASP}},
  title = {{LLM} Top 10 for Large Language Model Applications 2025},
  howpublished = {OWASP Gen AI Security Project},
  year = {2025},
  note = {Available: \url{https://owasp.org/www-project-top-10-for-large-language-model-applications/assets/PDF/OWASP-Top-10-for-LLMs-v2025.pdf}},
}

@techreport{ref15,
  author = {{NIST}},
  title = {{AI} Risk Management Framework ({AI} {RMF} 1.0)},
  institution = {National Institute of Standards and Technology},
  year = {2023},
  doi = {10.6028/NIST.AI.100-1},
  note = {Available: \url{https://nvlpubs.nist.gov/nistpubs/ai/NIST.AI.100-1.pdf}},
}

@misc{ref16,
  author = {{ServiceNow}},
  title = {{[Security Advisory]} {CVE-2025-12420}: Privilege Escalation via Prompt Injection in {ServiceNow} {AI} Platform ({BodySnatcher})},
  howpublished = {ServiceNow Support},
  year = {2026},
  month = jan,
  note = {Available: \url{https://support.servicenow.com/kb?id=kb_article_view\&sysparm_article=KB2587329}},
}

@misc{ref17,
  author = {{CSO Online}},
  title = {{AI} Disinformation Did Not Upend 2024 Elections, but the Threat Is Very Real},
  howpublished = {CSO Online},
  year = {2025},
  month = apr,
  note = {Available: \url{https://www.csoonline.com/article/3852770/}},
}

@misc{ref18,
  author = {{NIST Center for AI Standards and Innovation (CAISI)}},
  title = {{AI} Agent Standards Initiative},
  howpublished = {National Institute of Standards and Technology},
  year = {2026},
  month = feb,
  note = {Available: \url{https://www.nist.gov/caisi/ai-agent-standards-initiative}},
}

@techreport{ref19,
  author = {Galluzzo, R. and Fisher, B. and Booth, H. and Roberts, J.},
  title = {Accelerating the Adoption of Software and Artificial Intelligence Agent Identity and Authorization},
  institution = {NIST National Cybersecurity Center of Excellence},
  year = {2026},
  month = mar,
  type = {Draft Concept Paper},
  note = {Available: \url{https://www.nccoe.nist.gov/projects/software-and-ai-agent-identity-and-authorization}},
}

@inproceedings{ref20,
  author = {Coutinho, M. A. and Ashofteh, A. and {Al Helaly}, Y.},
  title = {Risk Taxonomies and Governance Frameworks for Generative {AI}: A Review of Ethical, Cybersecurity, and Regulatory Challenges},
  booktitle = {Proc. 20th Iberian Conf. Information Systems and Technologies (CISTI 2025)},
  series = {Lecture Notes in Networks and Systems},
  volume = {1717},
  publisher = {Springer},
  address = {Cham},
  year = {2026},
  doi = {10.1007/978-3-032-10721-3_1},
  note = {Available: \url{https://link.springer.com/chapter/10.1007/978-3-032-10721-3_1}},
}

@misc{ref21,
  author = {{Omada Identity}},
  title = {The State of Identity Governance Report 2026},
  howpublished = {Omada},
  year = {2026},
  month = jan,
  note = {Available: \url{https://omadaidentity.com/resources/analyst-reports/state-of-iga/}},
}

@article{ref22,
  author = {Moura, J. and others},
  title = {Prompt Injection Attacks in Large Language Models and {AI} Agent Systems},
  journal = {Information},
  volume = {17},
  number = {1},
  pages = {54},
  year = {2026},
  doi = {10.3390/info17010054},
}

@misc{ref23,
  author = {Ding, J. and others},
  title = {Spy vs. {AI}: How Artificial Intelligence Will Remake Espionage},
  howpublished = {Foreign Affairs},
  year = {2025},
  month = jan,
  note = {Available: \url{https://www.foreignaffairs.com/united-states/spy-vs-ai}},
}

@article{ref24,
  author = {{Springer AI and Ethics}},
  title = {Waging Warfare Against States: The Deployment of Artificial Intelligence in Cyber Espionage},
  journal = {AI and Ethics},
  year = {2025},
  doi = {10.1007/s43681-024-00628-x},
}

@inproceedings{ref25,
  author = {Rizvi, S. and Kurtz, A. and Pfeffer, J. and Rizvi, M.},
  title = {Securing the {Internet} of {Things} ({IoT}): A Security Taxonomy for {IoT}},
  booktitle = {Proc. 17th IEEE Int. Conf. Trust, Security and Privacy in Computing and Communications (TrustCom/BigDataSE 2018)},
  address = {New York, NY},
  year = {2018},
  month = aug,
  pages = {163--168},
  doi = {10.1109/TrustCom/BigDataSE.2018.00034},
  note = {Available: \url{https://ieeexplore.ieee.org/document/8455902}}
}

@techreport{ref26,
  author = {Rose, S. and Borchert, O. and Mitchell, S. and Connelly, S.},
  title = {Zero Trust Architecture},
  institution = {National Institute of Standards and Technology},
  number = {NIST Special Publication 800-207},
  year = {2020},
  month = aug,
  doi = {10.6028/NIST.SP.800-207},
}

@misc{ref27,
  author = {{OpenID Foundation}},
  title = {Identity Management for Agentic {AI}},
  howpublished = {OpenID Foundation Working Group Report},
  year = {2025},
  month = oct,
  note = {Available: \url{https://openid.net/wp-content/uploads/2025/10/Identity-Management-for-Agentic-AI.pdf}},
}

@misc{ref28,
  author = {Pandey, R.},
  title = {The Agentic {AI} Governance Framework: A Universal Model for Risk, Accountability, and Compliance in Autonomous Systems},
  howpublished = {SSRN},
  year = {2025},
  month = oct,
  doi = {10.2139/ssrn.5652350},
  note = {Available: \url{https://papers.ssrn.com/sol3/papers.cfm?abstract_id=5652350}},
}

@misc{ref29,
  author = {Joshi, S.},
  title = {Framework for Government Policy on Agentic and Generative {AI}: Governance, Regulation, and Risk Management},
  howpublished = {SSRN},
  year = {2025},
  month = aug,
  note = {Available: \url{https://papers.ssrn.com/sol3/papers.cfm?abstract_id=5511060}},
}

@misc{ref30,
  author = {Chun, Y. and others},
  title = {Comparative Global {AI} Regulation: Policy Perspectives from the {EU}, {China}, and the {US}},
  howpublished = {arXiv:2410.21279},
  year = {2024},
  month = oct,
  note = {Available: \url{https://arxiv.org/abs/2410.21279}},
}

@article{ref31,
  author = {Lim, L. S.},
  title = {Artificial Intelligence Regulation Matures: Landscapes of the {USA}, {European Union}, and {China}},
  journal = {Journal of Information Technology},
  year = {2025},
  doi = {10.1177/03400352251384915},
}

@misc{ref32,
  author = {{Communications of the ACM}},
  title = {Three Rulebooks, One Race: {AI} Regulation in the {U.S.}, {EU}, and {China}},
  howpublished = {Commun. ACM},
  year = {2026},
  month = feb,
  note = {Available: \url{https://cacm.acm.org/news/three-rulebooks-one-race-ai-regulation-in-the-u-s-eu-and-china/}},
}

@misc{ref33,
  author = {Dorwart, H. and Qu, H. and Brautigam, T. and Gong, J.},
  title = {Preparing for Compliance: Key Differences Between {EU} and Chinese {AI} Regulations},
  howpublished = {IAPP},
  year = {2025},
  month = feb,
  note = {Available: \url{https://iapp.org/news/a/preparing-for-compliance-key-differences-between-eu-chinese-ai-regulations}},
}

@article{ref34,
  author = {Deng, Z. and others},
  title = {{AI} Agents Under Threat: A Survey of Key Security Challenges and Future Pathways},
  journal = {ACM Comput. Surv.},
  volume = {57},
  number = {7},
  pages = {182},
  year = {2025},
  month = feb,
  doi = {10.1145/3716628},
}

@article{ref35,
  author = {He, Z. and others},
  title = {The Emerged Security and Privacy of {LLM} Agents: A Survey with Case Studies},
  journal = {ACM Comput. Surv.},
  year = {2025},
  doi = {10.1145/3773080},
}

@article{ref36,
  author = {Ferrag, M. A. and others},
  title = {From Prompt Injections to Protocol Exploits: Threats in {LLM}-Powered {AI} Agent Workflows},
  journal = {ICT Express},
  year = {2025},
  note = {Available: \url{https://www.sciencedirect.com/science/article/pii/S2405959525001997}},
}

@misc{ref37,
  author = {Castro, C. and others},
  title = {A Survey of Agentic {AI} and Cybersecurity: Challenges, Opportunities and Use-Case Prototypes},
  howpublished = {arXiv:2601.05293},
  year = {2026},
  month = jan,
  note = {Available: \url{https://arxiv.org/html/2601.05293v1}},
}

@misc{ref38,
  author = {Shavit, Y. and Agarwal, S. and others},
  title = {Practices for Governing Agentic {AI} Systems},
  howpublished = {OpenAI},
  year = {2023},
  month = dec,
  note = {Available: \url{https://cdn.openai.com/papers/practices-for-governing-agentic-ai-systems.pdf}},
}

@misc{ref39,
  author = {Holgersson, M. and Dahlander, L. and Chesbrough, H. W. and Bogers, M.},
  title = {Rethinking {AI} Agents: A Principal-Agent Perspective},
  howpublished = {California Management Review},
  year = {2025},
  month = jul,
  note = {Available: \url{https://cmr.berkeley.edu/2025/07/rethinking-ai-agents-a-principal-agent-perspective/}},
}

@article{ref40,
  author = {Kaur, D. and others},
  title = {{AI} Governance: A Systematic Literature Review},
  journal = {AI and Ethics},
  year = {2025},
  month = jan,
  doi = {10.1007/s43681-024-00653-w},
}

@article{ref41,
  author = {Csernatoni, R.},
  title = {Global {AI} Governance: Barriers and Pathways Forward},
  journal = {International Affairs},
  volume = {100},
  number = {3},
  pages = {1275--1294},
  year = {2024},
  month = may,
  doi = {10.1093/ia/iiae084},
}

@article{ref42,
  author = {others},
  title = {Navigating the Nexus: Geopolitical, International Relations and Technical Dimensions of {US}-{China} Cyber Strategic Competition},
  journal = {Cogent Social Sciences},
  year = {2025},
  month = may,
  doi = {10.1080/23311886.2025.2499171},
}

@misc{ref43,
  author = {Pouget, H. and Dennis, C. and others},
  title = {The Future of International Scientific Assessments of {AI}'s Risks},
  howpublished = {Carnegie Endowment for International Peace},
  year = {2024},
  month = aug,
  note = {Available: \url{https://carnegieendowment.org/research/2024/08/the-future-of-international-scientific-assessments-of-ais-risks}},
}

@misc{ref44,
  author = {{NIST NCCoE}},
  title = {Software and {AI} Agent Identity and Authorization},
  howpublished = {National Cybersecurity Center of Excellence},
  year = {2026},
  note = {Available: \url{https://www.nccoe.nist.gov/projects/software-and-ai-agent-identity-and-authorization}},
}

@techreport{ref45,
  author = {{Entro Labs}},
  title = {{NHI} and Secrets Risk Report --- {H1} 2025},
  institution = {Entro Security},
  year = {2025},
  month = jul,
  note = {Available: \url{https://www.cybersecuritytribe.com/news/research-reveals-44-growth-in-nhis-from-2024-to-2025}},
}

@misc{ref46,
  author = {{European Commission}},
  title = {{EU} {AI} Act},
  howpublished = {Official Journal of the European Union},
  year = {2024},
  month = jul,
  note = {Available: \url{https://digital-strategy.ec.europa.eu/en/policies/regulatory-framework-ai}},
}

@misc{ref47,
  author = {{EU AI Act Service Desk}},
  title = {Timeline for the Implementation of the {EU} {AI} Act},
  howpublished = {European Commission},
  year = {2025},
  note = {Available: \url{https://ai-act-service-desk.ec.europa.eu/en/ai-act/timeline/timeline-implementation-eu-ai-act}},
}

@misc{ref48,
  author = {{Pearl Cohen}},
  title = {New Guidance Under the {EU} {AI} Act Ahead of Its Next Enforcement Date},
  howpublished = {Pearl Cohen Law},
  year = {2025},
  month = dec,
  note = {Available: \url{https://www.pearlcohen.com/new-guidance-under-the-eu-ai-act-ahead-of-its-next-enforcement-date/}},
}

@misc{ref49,
  author = {{K\&L Gates}},
  title = {{EU} and {Luxembourg} Update on the {European} Harmonised Rules on Artificial Intelligence},
  howpublished = {K\&L Gates},
  year = {2026},
  month = jan,
  note = {Available: \url{https://www.klgates.com/EU-and-Luxembourg-Update-on-the-European-Harmonised-Rules-on-Artificial-IntelligenceRecent-Developments-1-20-2026}},
}

@misc{ref50,
  author = {{Executive Office of the President}},
  title = {Ensuring a National Policy Framework for Artificial Intelligence},
  howpublished = {Executive Order 14365, Federal Register},
  year = {2025},
  month = dec,
  note = {Available: \url{https://www.federalregister.gov/documents/2025/12/16/2025-23092/ensuring-a-national-policy-framework-for-artificial-intelligence}},
}

@misc{ref51,
  author = {{White House Office of Science and Technology Policy}},
  title = {Winning the Race: {America}'s {AI} Action Plan},
  howpublished = {The White House},
  year = {2025},
  month = jul,
  note = {Available: \url{https://www.whitehouse.gov/wp-content/uploads/2025/07/Americas-AI-Action-Plan.pdf}},
}

@misc{ref52,
  author = {{Baker Botts}},
  title = {{U.S.} Artificial Intelligence Law Update: Navigating the Evolving State and Federal Regulatory Landscape},
  howpublished = {Baker Botts},
  year = {2026},
  month = jan,
  note = {Available: \url{https://www.bakerbotts.com/thought-leadership/publications/2026/january/us-ai-law-update}},
}

@misc{ref53,
  author = {Chen, H.-Y.},
  title = {{AI} Governance and Regulation 2026: A Complete Guide to Global Frameworks},
  year = {2026},
  month = feb,
  note = {Available: \url{https://www.hungyichen.com/en/insights/ai-governance-regulatory-landscape-2026}},
}

@misc{ref54,
  author = {{Linklaters}},
  title = {China's 2025 Cybersecurity Law Amendments: Enhanced Penalties, Expanded Extraterritorial Application, and {AI} Governance},
  howpublished = {Linklaters Tech Insights},
  year = {2025},
  month = oct,
  note = {Available: \url{https://techinsights.linklaters.com/post/102lrz5/}},
}

@misc{ref55,
  author = {{IAPP}},
  title = {Global {AI} Governance Law and Policy: {China}},
  howpublished = {International Association of Privacy Professionals},
  year = {2025},
  note = {Available: \url{https://iapp.org/resources/article/global-ai-governance-china}},
}

@misc{ref56,
  author = {{ICLG}},
  title = {China's Key Developments in Artificial Intelligence Governance in 2025},
  howpublished = {ICLG},
  year = {2025},
  month = dec,
  note = {Available: \url{https://iclg.com/practice-areas/telecoms-media-and-internet-laws-and-regulations/03-china-s-key-developments-in-artificial-intelligence-governance-in-2025}},
}

@misc{ref57,
  author = {{ANSI}},
  title = {China Announces Action Plan for Global {AI} Governance},
  howpublished = {American National Standards Institute},
  year = {2025},
  month = aug,
  note = {Available: \url{https://www.ansi.org/standards-news/all-news/8-1-25-china-announces-action-plan-for-global-ai-governance}},
}

@misc{ref58,
  author = {{Microsoft Threat Intelligence}},
  title = {Silk Typhoon Targeting {IT} Supply Chain},
  howpublished = {Microsoft Security Blog},
  year = {2025},
  month = mar,
  note = {Available: \url{https://www.microsoft.com/en-us/security/blog/2025/03/05/silk-typhoon-targeting-it-supply-chain/}},
}

@misc{ref59,
  author = {Johnson, D. B.},
  title = {Silk Typhoon Shifted to Specifically Targeting {IT} Management Companies},
  howpublished = {CyberScoop},
  year = {2025},
  month = mar,
  note = {Available: \url{https://cyberscoop.com/silk-typhoon-targets-it-services/}},
}

@misc{ref60,
  author = {{Federal Bureau of Investigation}},
  title = {{FBI} Announces Joint Cybersecurity Advisory Related to {Salt} Typhoon},
  howpublished = {FBI.gov},
  year = {2025},
  month = aug,
  note = {Available: \url{https://www.fbi.gov/video-repository/salttyphoon082725.mp4/view}},
}

@techreport{ref61,
  author = {{Congressional Research Service}},
  title = {Salt Typhoon Hacks of Telecommunications Companies and Federal Response Implications},
  institution = {Congress.gov},
  year = {2025},
  month = jan,
  note = {Available: \url{https://www.congress.gov/crs-product/IF12798}},
}

@misc{ref62,
  author = {Johnson, D. B.},
  title = {Officials Worry Salt Typhoon Apathy Is Killing Momentum for Tougher Telecom Security Rules},
  howpublished = {CyberScoop},
  year = {2026},
  month = mar,
  note = {Available: \url{https://cyberscoop.com/salt-typhoon-china-telecom-hack-impact-new-jersey/}},
}

@misc{ref63,
  author = {Huang, K. and Narajala, V. S. and Yeoh, J. and Raskar, R. and others},
  title = {A Novel Zero-Trust Identity Framework for Agentic {AI}: Decentralized Authentication and Fine-Grained Access Control},
  howpublished = {arXiv:2505.19301},
  year = {2025},
  month = may,
  note = {Available: \url{https://arxiv.org/abs/2505.19301}},
}

@misc{ref64,
  author = {{MassiveScale.AI / Cloud Security Alliance}},
  title = {The Agentic Trust Framework: Zero Trust Governance for {AI} Agents},
  howpublished = {CSA Blog},
  year = {2026},
  month = feb,
  note = {Available: \url{https://cloudsecurityalliance.org/blog/2026/02/02/the-agentic-trust-framework-zero-trust-governance-for-ai-agents}},
}

@misc{ref65,
  author = {{OWASP GenAI Security Project}},
  title = {{OWASP} Top 10 for Agentic Applications 2026},
  howpublished = {OWASP},
  year = {2025},
  month = dec,
  note = {Available: \url{https://genai.owasp.org/resource/owasp-top-10-for-agentic-applications-for-2026/}},
}

@misc{ref66,
  author = {Gupta, V.},
  title = {The Looming Authorization Crisis {IAM}},
  howpublished = {ISACA Journal},
  volume = {5},
  year = {2025},
  month = sep,
  note = {Available: \url{https://www.isaca.org/resources/news-and-trends/industry-news/2025/the-looming-authorization-crisis-why-traditional-iam-fails-agentic-ai}},
}

@misc{ref67,
  author = {{W3C}},
  title = {Decentralized Identifiers ({DIDs}) v1.0},
  howpublished = {W3C Recommendation},
  year = {2022},
  month = jul,
  note = {Available: \url{https://www.w3.org/TR/did-core/}},
}

@misc{ref68,
  author = {{Cloud Native Computing Foundation}},
  title = {{SPIFFE}: Secure Production Identity Framework for Everyone},
  howpublished = {CNCF Graduated Project},
  year = {2024},
  note = {Available: \url{https://spiffe.io/}},
}

@misc{ref69,
  author = {{HashiCorp}},
  title = {{SPIFFE}: Securing the Identity of Agentic {AI} and Non-Human Actors},
  howpublished = {HashiCorp Blog},
  year = {2025},
  note = {Available: \url{https://www.hashicorp.com/en/blog/spiffe-securing-the-identity-of-agentic-ai-and-non-human-actors}},
}

@misc{ref70,
  author = {{Optimum Partners}},
  title = {{AI} Security Architecture: Implementing Workload Identity Federation ({WIF}) and {SPIFFE}},
  year = {2026},
  month = jan,
  note = {Available: \url{https://optimumpartners.com/insight/ai-security-architecture-implementing-workload-identity-federation-wif-and-spiffe/}},
}

@misc{ref71,
  author = {{Lumos}},
  title = {Agentic {AI} and Identity Governance: What You Need to Know},
  year = {2025},
  month = dec,
  note = {Available: \url{https://www.lumos.com/topic/agentic-ai-identity-governance-management}},
}

@misc{ref72,
  author = {{Cisco Talos}},
  title = {2024 Year in Review},
  howpublished = {Cisco Systems},
  year = {2025},
  note = {Available: \url{https://blog.talosintelligence.com/2024yearinreview/}},
}

@misc{ref73,
  author = {Alkove, J. and others},
  title = {Practitioner's Guide to the Future of Identity: New Maturity Model},
  howpublished = {Identity Working Group, Oleria},
  year = {2025},
  note = {Available: \url{https://www.oleria.com/resources/guides/nhi}},
}

@misc{ref74,
  author = {{Permiso Security}},
  title = {What Are Non-Human Identities? {Complete} Guide to {NHI} Security for 2025},
  howpublished = {Permiso},
  year = {2025},
  note = {Available: \url{https://permiso.io/non-human-identity-nhi-security-guide}},
}

@misc{ref75,
  author = {{KuppingerCole Analysts}},
  title = {Mastering Non-Human Identity Governance for Enhanced Security and Efficiency},
  howpublished = {KuppingerCole Blog},
  year = {2025},
  month = oct,
  note = {Available: \url{https://www.kuppingercole.com/blog/kuppinger/mastering-non-human-identity-governance-for-enhanced-security-and-efficiency}},
}

@misc{ref76,
  author = {{OASIS Security}},
  title = {Non-Human Identity Management Guide: Audit, Certify and Govern {NHIs}},
  howpublished = {OASIS Security Blog},
  year = {2026},
  month = jan,
  note = {Available: \url{https://www.oasis.security/blog/non-human-identity-management}},
}

@misc{ref77,
  author = {Ozkaya, E.},
  title = {Enterprise {AI} Security and Governance Roadmap (2026 {CISO} Strategy)},
  year = {2026},
  month = feb,
  note = {Available: \url{https://erdalozkaya.com/enterprise-ai-security/}},
}

@misc{ref78,
  author = {{CyberArk}},
  title = {2025 Identity Security Landscape},
  howpublished = {CyberArk},
  year = {2025},
  month = apr,
  note = {Available: \url{https://www.cyberark.com/press/machine-identities-outnumber-humans-by-more-than-80-to-1-new-report-exposes-the-exponential-threats-of-fragmented-identity-security/}},
}

@misc{ref79,
  author = {{AppViewX}},
  title = {Key Takeaways from the 2024 {ESG} Report on Non-Human Identity ({NHI}) Management},
  howpublished = {AppViewX Blog},
  year = {2024},
  month = oct,
  note = {Available: \url{https://www.appviewx.com/blogs/key-takeaways-from-the-2024-esg-report-on-non-human-identity-nhi-management/}},
}

@misc{ref80,
  author = {Markovic, S.},
  title = {Non-Human Identities Push Identity Security into Uncharted Territory},
  howpublished = {Help Net Security},
  year = {2025},
  month = dec,
  note = {Available: \url{https://www.helpnetsecurity.com/2025/12/30/identity-security-permissions-sprawl/}},
}

@misc{ref81,
  author = {{Research and Markets}},
  title = {Non-Human Identity Solutions, Global, 2024--2030},
  howpublished = {Research and Markets},
  year = {2026},
  month = feb,
  note = {Available: \url{https://www.researchandmarkets.com/reports/6217813/non-human-identity-solutions-global}},
}

@misc{ref82,
  author = {Gupta, D.},
  title = {The {AI} Identity Crisis: {NHIs}, Agents and Vibe Coding in 2025},
  year = {2025},
  month = oct,
  note = {Available: \url{https://guptadeepak.com/the-identity-crisis-no-ones-talking-about-how-ai-agents-and-vibe-coding-are-rewriting-the-rules-of-digital-security/}},
}

@misc{ref83,
  author = {{Tufin}},
  title = {The Lasting Impact of the {CrowdStrike} Update Outage},
  howpublished = {Tufin Blog},
  year = {2025},
  month = jun,
  note = {Available: \url{https://www.tufin.com/blog/lasting-impact-of-crowdstrike-update-outage}},
}

@misc{ref84,
  author = {Duffy, C.},
  title = {{CrowdStrike} Outage: We Finally Know What Caused It and How Much It Cost},
  howpublished = {CNN Business},
  year = {2024},
  month = jul,
  note = {Available: \url{https://www.cnn.com/2024/07/24/tech/crowdstrike-outage-cost-cause}},
}

@misc{ref85,
  author = {Snider, S.},
  title = {{CrowdStrike} Outage Drained \$5.4 Billion from {Fortune} 500: Report},
  howpublished = {InformationWeek},
  year = {2024},
  month = jul,
  note = {Available: \url{https://www.informationweek.com/cyber-resilience/crowdstrike-outage-drained-5-4-billion-from-fortune-500-report}},
}

@misc{ref86,
  author = {{Messageware}},
  title = {What Caused the {CrowdStrike} Outage: A Detailed Breakdown},
  year = {2024},
  month = aug,
  note = {Available: \url{https://www.messageware.com/what-caused-the-crowdstrike-outage-a-detailed-breakdown/}},
}

@misc{ref87,
  author = {{Reuters}},
  title = {Delta Can Sue {CrowdStrike} Over Computer Outage That Caused 7,000 Canceled Flights},
  howpublished = {Reuters},
  year = {2025},
  month = may,
  note = {Available: \url{https://www.reuters.com/sustainability/boards-policy-regulation/delta-can-sue-crowdstrike-over-computer-outage-that-caused-7000-canceled-flights-2025-05-19/}},
}

@misc{ref88,
  author = {{Harvard Business Review}},
  title = {What the 2024 {CrowdStrike} Glitch Can Teach Us About Cyber Risk},
  howpublished = {HBR},
  year = {2025},
  month = jan,
  note = {Available: \url{https://hbr.org/2025/01/what-the-2024-crowdstrike-glitch-can-teach-us-about-cyber-risk}},
}

@misc{ref89,
  author = {{Gravitee.io}},
  title = {State of {AI} Agent Security 2026 Report: When Adoption Outpaces Control},
  year = {2026},
  month = feb,
  note = {Available: \url{https://www.gravitee.io/blog/state-of-ai-agent-security-2026-report-when-adoption-outpaces-control}},
}

@misc{ref90,
  author = {{Strata Identity / Cloud Security Alliance}},
  title = {The {AI} Agent Identity Crisis: New Research Reveals a Governance Gap},
  year = {2026},
  month = feb,
  note = {Available: \url{https://www.strata.io/blog/agentic-identity/the-ai-agent-identity-crisis-new-research-reveals-a-governance-gap/}},
}

@misc{ref91,
  author = {Chmura, M.},
  title = {Agentic {AI}: The Future and Governance of Autonomous Systems},
  howpublished = {Bloomsbury Intelligence and Security Institute},
  year = {2026},
  month = feb,
  note = {Available: \url{https://bisi.org.uk/reports/agentic-ai-the-future-and-governance-of-autonomous-systems}},
}

@misc{ref92,
  author = {{Lumos}},
  title = {New Research Finds Identity has Become the Most Common Entry Point for Cyberattacks},
  year = {2026},
  month = feb,
  note = {Available: \url{https://www.prnewswire.com/news-releases/new-research-finds-identity-has-become-the-most-common-entry-point-for-cyberattacks-302695637.html}},
}

@book{ref93,
  editor = {Schmitt, M. N.},
  title = {Tallinn Manual 2.0 on the International Law Applicable to Cyber Operations},
  edition = {2nd},
  publisher = {Cambridge University Press},
  address = {Cambridge, UK},
  year = {2017},
  doi = {10.1017/9781316822524},
}

@misc{ref94,
  author = {{IANS Research}},
  title = {{AI} Agents Are Creating an Identity Security Crisis in 2026},
  howpublished = {IANS Research Blog},
  year = {2026},
  month = feb,
  note = {Available: \url{https://www.iansresearch.com/resources/all-blogs/post/security-blog/2026/02/24/ai-agents-are-creating-an-identity-security-crisis-in-2026}},
}

@article{ref95,
  author = {Perboli, G. and Simionato, N. and Pratali, S.},
  title = {Navigating the {AI} Regulatory Landscape: Balancing Innovation, Ethics, and Global Governance},
  journal = {Economic and Political Studies},
  publisher = {Taylor and Francis},
  volume = {13},
  number = {4},
  pages = {367--397},
  year = {2025},
  month = dec,
  doi = {10.1080/20954816.2025.2569584},
}

@article{ref96,
  author = {Schmitt, L.},
  title = {Mapping Global {AI} Governance: A Nascent Regime in a Fragmented Landscape},
  journal = {AI and Ethics},
  volume = {2},
  pages = {303--314},
  year = {2022},
  doi = {10.1007/s43681-021-00083-y},
}

@misc{ref97,
  author = {Zorz, Z.},
  title = {Marks and Spencer Ransomware Breach Incident},
  howpublished = {Help Net Security},
  year = {2025},
  month = apr,
  note = {Available: \url{https://www.helpnetsecurity.com/2025/04/29/marks-spencer-ransomware-breach-incident/}},
}

@misc{ref98,
  author = {{National Crime Agency}},
  title = {{NCA} Retail cyber attacks: NCA arrest four for attacks on M\&S, Co-op and Harrods},
  howpublished = {NCA},
  year = {2025},
  month = jul,
  note = {Available: \url{https://www.nationalcrimeagency.gov.uk/news/retail-cyber-attacks-nca-arrest-four-for-attacks-on-m-s-co-op-and-harrods}},
}

@article{ref99,
  author = {Wang, S. and Zhang, Y. and Xiao, Y. and Liang, Z.},
  title = {Artificial Intelligence Policy Frameworks in {China}, the {European Union} and {United States}: An Analysis Based on Structure Topic Model},
  journal = {Technological Forecasting and Social Change},
  volume = {212},
  pages = {123971},
  year = {2025},
  doi = {10.1016/j.techfore.2025.123971},
}

@inproceedings{krawiecka2016protecting,
  author    = {Krawiecka, Klaudia and Paverd, Andrew and Asokan, N.},
  title     = {Protecting Password Databases Using Trusted Hardware},
  booktitle = {Proceedings of the 1st Workshop on System Software 
               for Trusted Execution (SysTEX '16)},
  year      = {2016},
  publisher = {ACM},
  pages     = {1--6},
  doi       = {10.1145/3007788.3007798}
}

@article{bodea2025trusted,
  author    = {Bodea, Teofil and Misono, Masanori and Pritzi, Julian 
               and Sabanic, Patrick and Sommer, Thore 
               and Unnibhavi, Harshavardhan and Schall, David 
               and Santos, Nuno and Stavrakakis, Dimitrios 
               and Bhatotia, Pramod},
  title     = {Trusted {AI} Agents in the Cloud},
  journal   = {arXiv preprint arXiv:2512.05951},
  year      = {2025},
  url       = {https://arxiv.org/abs/2512.05951}
}

@inproceedings{rodriguezgarzon2026agents,
  author    = {Rodriguez Garzon, Sandro and Vaziry, Awid 
               and Kuzu, Enis Mert and Gehrmann, Dennis Enrique 
               and Varkan, Buse and Gaballa, Alexander 
               and K{\"u}pper, Axel},
  title     = {{AI} Agents with Decentralized Identifiers 
               and Verifiable Credentials},
  booktitle = {Proceedings of the 18th International Conference 
               on Agents and Artificial Intelligence (ICAART 2026)},
  year      = {2026},
  volume    = {1},
  pages     = {252--259},
  publisher = {SCITEPRESS},
  doi       = {10.48550/arXiv.2511.02841}
}

@techreport{ietf2023rats,
  author      = {Birkholz, Henk and Thaler, Dave and Richardson, Michael 
                 and Smith, Ned and Pan, Wei},
  title       = {Remote Attestation Procedures ({RATS}) Architecture},
  institution = {Internet Engineering Task Force},
  year        = {2023},
  type        = {RFC},
  number      = {9334},
  url         = {https://www.rfc-editor.org/rfc/rfc9334}
}

@article{cheng2024intel,
  author    = {Cheng, Pau-Chen and Ozga, Wojciech and Valdez, Enriquillo 
               and Ahmed, Salman and Gu, Zhongshu and Jamjoom, Hani 
               and Franke, Hubertus and Bottomley, James},
  title     = {Intel {TDX} Demystified: A Top-Down Approach},
  journal   = {ACM Computing Surveys},
  year      = {2024},
  volume    = {56},
  number    = {9},
  articleno = {238},
  pages     = {1--33},
  doi       = {10.1145/3652597}
}

@misc{costan2016intel,
  author       = {Costan, Victor and Devadas, Srinivas},
  title        = {Intel {SGX} Explained},
  year         = {2016},
  howpublished = {IACR Cryptology ePrint Archive, Report 2016/086},
  url          = {https://eprint.iacr.org/2016/086}
}

@book{arthur2015practical,
  author    = {Arthur, Will and Challener, David},
  title     = {A Practical Guide to {TPM} 2.0: Using the Trusted Platform 
               Module in the New Age of Security},
  year      = {2015},
  publisher = {Apress},
  doi       = {10.1007/978-1-4302-6584-9}
}

@misc{hidglobal2025trust,
  author       = {{HID Global}},
  title        = {Trust Standards Evolve: {AI} Agents, the Next Chapter 
                  for {PKI}},
  year         = {2025},
  month        = {November},
  howpublished = {HID Global Blog},
  url          = {https://blog.hidglobal.com/trust-standards-evolve-ai-agents-next-chapter-pki}
}

@misc{castro2026survey,
  author       = {Castro, Carlos and others},
  title        = {A Survey of Agentic {AI} and Cybersecurity: 
                  Challenges, Opportunities and Use-Case Prototypes},
  year         = {2026},
  howpublished = {arXiv preprint arXiv:2601.05293},
  url          = {https://arxiv.org/abs/2601.05293}
}

@misc{databricks2026dasf,
  author       = {{Databricks}},
  title        = {Agentic {AI} Security: New Risks and Controls 
                  in the Databricks {AI} Security Framework ({DASF} v3.0)},
  year         = {2026},
  month        = {March},
  howpublished = {Databricks Blog},
  url          = {https://www.databricks.com/blog/agentic-ai-security-new-risks-and-controls-databricks-ai-security-framework-dasf-v30}
}

@article{kumar2025nextgen,
  author    = {Kumar, Pankaj},
  title     = {Next-Generation Secure Authentication and Access Control 
               Architectures: Advanced Techniques for Securing Distributed 
               Systems in Modern Enterprises},
  journal   = {International Journal of Computational and Experimental 
               Science and Engineering},
  year      = {2025},
  volume    = {11},
  number    = {3},
  doi       = {10.22399/ijcesen.3294}
}

@misc{obsidian2026landscape,
  author       = {{Obsidian Security}},
  title        = {The 2025 {AI} Agent Security Landscape: 
                  Players, Trends, and Risks},
  year         = {2026},
  month        = {January},
  howpublished = {Obsidian Security Blog},
  url          = {https://www.obsidiansecurity.com/blog/ai-agent-market-landscape}
}

@misc{owasp2025agentic,
  author       = {{OWASP GenAI Security Project}},
  title        = {{OWASP} Top 10 for Agentic Applications 2026},
  year         = {2025},
  month        = {December},
  howpublished = {OWASP},
  url          = {https://genai.owasp.org/resource/owasp-top-10-for-agentic-applications-for-2026/}
}

\end{document}